\documentclass[12pt,a4paper,notitlepage]{article}
\usepackage[utf8]{inputenc}
\usepackage[english]{babel}
\usepackage[T1]{fontenc}
\usepackage{hyperref}
\usepackage{amsmath}
\usepackage{amsfonts}
\usepackage{amssymb}
\usepackage{color} 
\usepackage{caption}
\usepackage{array,multirow,makecell}
\setcellgapes{1pt}
\makegapedcells
\newcolumntype{R}[1]{>{\raggedleft\arraybackslash }b{#1}}
\newcolumntype{L}[1]{>{\raggedright\arraybackslash }b{#1}}
\newcolumntype{C}[1]{>{\centering\arraybackslash }b{#1}}
\newcommand{\Tr}{\mathrm{Tr}}

\newtheorem{theorem}{Theorem}
\newtheorem{definition}{Definition}
\newtheorem{proposition}{Proposition}
\newtheorem{remark}{Remark}
\newtheorem{claim}{Claim}
\newtheorem{corollary}{Corollary}

\newcommand{\sym}{\mathrm{Sym}}
\newcommand{\perm}{\mathrm{perm}}

\usepackage{enumerate}
\usepackage{subfig}

\newcommand{\mZ}{{\mathbb Z}}
\newcommand{\vp}{{\varphi}}
\newcommand{\bvp}{{\bar{\varphi}}}
\newcommand{\beq}{\begin{equation}}
\newcommand{\eeq}{\end{equation}}
\newcommand{\bea}{\begin{eqnarray}}
\newcommand{\eea}{\end{eqnarray}}
\definecolor{mygray}{gray}{0.3}

\newcommand{\bes}{\begin{eqnarray}}
\newcommand{\ees}{\end{eqnarray}}

\newcommand\restr[2]{{
  \left.\kern-\nulldelimiterspace 
  #1 
  \vphantom{\big|} 
  \right|_{#2} 
  }}

\usepackage{multicol}
\usepackage{amssymb}
\usepackage{graphicx}

\makeatletter
\begin{document}
\begin{center}
\textbf{\Large{
Unitary symmetry constraints on tensorial group field theory
 renormalization group flow}}
\vspace{15pt}

{\large Vincent Lahoche$^a$\footnote{vincent.lahoche@th.u-psud.fr} and Dine Ousmane Samary$^{b,c}$\footnote{dine.ousmane.samary@aei.mpg.de}}
\vspace{0.5cm}

a)\, Laboratoire de Physique Th\'eorique, CNRS-UMR 8627, Universit\'e Paris-Sud 11, 91405 Orsay Cedex, France

b)\, Max Planck Institute for Gravitational Physics, Albert Einstein Institute, Am M\"uhlenberg 1, 14476, Potsdam, Germany

c)\,  Facult\'e des Sciences et Techniques/ ICMPA-UNESCO Chair, Universit\'e d'Abomey-
Calavi, 072 BP 50, Benin


\end{center}

\vspace{5pt}
\begin{abstract}
\noindent
Renormalization group    methods are an essential ingredient in the study of nonperturbative problems  of quantum field theory.  
This paper deal with the symmetry constraints on the renormalization group flow for quartic melonic tensorial group field theories. Using the  unitary invariance of the interactions, we provide a  set of Ward-Takahashi identities which leads to relations between correlation functions.  There are numerous reasons to consider such Ward identities in the functional renormalization group. Their compatibility along the flow provides  a non-trivial constraint on the reliability of the approximation schemes used in the non-perturbative regime, especially on the truncation and the choice of the regulator.  We establish the so called structure equations in the melonic sector and in the symmetric phase. As an example we consider the  $T^4_5$ TGFT model without gauge constraint. The Wetterich flow equation  is given and the way to  improve the  truncation on the effective action is also scrutinized.

\end{abstract}

\setcounter{tocdepth}{2}

\section{Introduction}
The construction of a quantum theory of gravity (QG), combining general relativity (GR) and quantum mechanics (QM),  may give a deeper understanding of the nature of space, time and geometry. QG become the tools for addressing questions about the origin of our universe, whose current description is incomplete due to the breakdown of GR at the Big Bang \cite{Oriti:2006ar}-\cite{Rovelli:2010bf}. In the last years there have been key developments in several approaches to quantum gravity, turning them into mature fields.
 Our approach in these directions is called tensorial group field theory (TGFT). It  is  a new class of field theories which aims to combine group field theories (GFTs) and tensor models (TMs), to enjoy renormalization and  asymptotic freedom in quite some generality, and for which we can show  another case of its coexistence with a Wilson–Fisher fixed point.   GFTs are quantum field theories over the group manifolds and are characterised by a specific form of non-locality in their interactions, with the basis variable being a complex field, function of $d$-group elements \cite{Oriti:2014yla}. It can be represented graphically as a $(d-1)$-simplex with field arguments associated to its faces, or as a $d$-valent graph vertex, with the field argument associated to its links. TMs generalize matrix models and are considered as a convenient formalism for studying random geometries \cite{Gurau:2009tw}-\cite{Rivasseau:2012yp}. Recently, these TMs have acknowledged a strong revival thanks to the discovery by Gurau of the analogue of the t’Hooft $1/N$-expansion for the tensor situation \cite{Gurau:2011xq}-\cite{Gurau:2010ba}  and of tensor renormalizable actions \cite{Carrozza:2012uv}-\cite{Geloun:2011cy}. The tensor model framework begins to take a growing role in the problem of quantum gravity and rises as a true alternative to several known approaches. 
The renormalizability of large class of TGFTs models ensure the quantum consistency  at macroscopic scales. On the other hand the computation of the nonperturbative renormalization group (RG) flow  to large scale  of the same models maybe help to identify the macroscopic structure  and probably show if the condensate phase exists see \cite{Gielen:2014uga}-\cite{Oriti:2005tx} and references therein.

The RG equations aim for a piecewise solution of the fluctuation problem. They describe the scale dependence of some type of “effective action”. 
It is   a nonperturbative method which allows us to interpolate smoothly between the UV laws and the IR phenomena in physical systems \cite{Wilson:1971bg}-\cite{Wilson:1971dh}. It realization to quantum field and statistical  theories  is called the functional renormalization group (FRG). FRG can be roughly described as a flow in a certain infinite dimensional functional space for actions, the theory space and have allowed to prove particularly the asymptotic freedom and the non-trivial IR behavior of the certain models in field theory. The main advantages of such a formulation are its flexibility when it comes to truncations of the full theory, as well as its numerical accessibility.  One truncates the infinite tower of flow equations for the $n$-point functions considering only vertices up to a given number of legs, possibly using various ansatzs for some of them \cite{Wetterich:1989xg}. 
Flows for a general class of correlation functions are derived, and it is shown how symmetry relations of the underlying theory are lifted to the regularised theory. The flow equation allows us to calculate the full effective action $\Gamma$ from an initial effective action $\Gamma_\Lambda$ if the latter is well under control ($\Lambda$ is some initial scale also called the UV cutoff).

Recently the analysis of the FRG to various TGFT models is performed \cite{Carrozza:2014rba}-\cite{Lahoche:2016xiq}. The occurence of non-perturbative fixed points and their critical behavior in the UV and IR is studied.  The confirmation of asymptotically freedom and safety is also given. In the first time  the simplest available truncation is used, and  in which only the perturbative relevant coupling constants are taken into account. In practice simple truncations $\Gamma=S$ usually detect quite easily non-trivial fixed points. In the other hand the truncation is extended to take into account the interaction of high order melonic contributions.  The existence of other fixed point can be proved rigorously and  make these extensions excellent approximations.  Despite the fact that very encouraging results have been obtained,  the question  with the consistency of the truncations remain unsatisfactory due to the lack of convergence in the flow. Also the choice of the regulator in the Wetterich equation  are given without prove of it consistency.  

The aims of the following  paper is to scrutinize the FRG in detail  for a TGFT models by interesting to alternative shemes. The Ward-Takahashi (WT) identities is used to defined the nontrivial constraint on the  flows. This leads to define the hyper-line in which the regulator can be chosen without clumsiness and also impose the chosen of the truncation in appropriate way. In the case of symmetric phase  a  nontrivial UV attractive fixed point is given. The numerical computation of the flow diagram is also set up. Note that our approach here is completely different from the usual FRG method. It is based on solving order by order the flows by using the WT-identities for $n$-point correlators as a constraint. 

This paper is organized as follows. In section \eqref{section2} we provide   the definition of the model and its symmetries. The FRG method is also given for particular $U(1)$-TGFT.  The symmetric and non symetric phase is discussed in section \eqref{section3}, in which we point out the melonic structure of the corresponding Feynman graphs. The recursive construction of the vacuum melons is also given.
The  section \eqref{section4} is devoted to  the WT-identities which result from the symmetry of the functional action.
The case of the symmetric phase is scrutinized. In section \eqref{section5} the melonic structure equations are given. We also provide the generalization of these equations respect to higher order melons which results from the gluing of elementary melon.  The exact melonic flow equations for local interactions is derived in the unitary symmetry constrained.  In section \eqref{section6} we give the way to improve the local truncation by using the structure equations. The conclusion and remarks are given in section \eqref{section7}.

\section{ Renormalization group flow for TGFT}\label{section2}
\subsection{A TGFT model without closure constraint}
We recall the main steps of the procedure leading to the FRG for  TGFT models described in       \cite{Benedetti:2015yaa}. The applications of FRG method to TGFT   leads to  the subtraction of all the divergences occurring in the perturbative expansion. The fact that a divergence occurs in the Ward identities for the initial value of the effective action, the TGFT models requires renormalization and the  choice of the initial conditions seems to be crucial in contrast with ordinary field theories. 
We define the TGFT model with the action $S[\vp,\bvp]$ in which the fields $\vp$ and it conjugate $\bvp$ take values on $d$-copies  on the Lie group $U(1)$ as
\bea
\vp: U(1)^d&&\longrightarrow \mathbb{C}\cr
\vec g:=(g_1,\cdots,g_d)&&\longmapsto\vp(\vec g)
\eea
such that $S[\vp,\bvp]$ takes the form
\bea
S[\vp,\bvp]=\int_{G} d\vec g\, \bvp(\vec g)(-\Delta+m^2)\vp(\vec g)+ S_{int}[\vp,\bvp]\,, 
\eea
where $\Delta$ is the sum of the Laplace-Beltrami operator on $U(1)$ i.e.
$
\Delta=\sum_{k=1}^d\Delta_k.
$
The interaction part of the action $S_{int}$ involves higher power of the fields, and for tensorial models, it is a sum of \textit{connected tensorial invariants}. A tensorial invariant is made with an equal number of $\vp$ and $\bar{\vp}$, whose arguments are identified and summed only between a $\vp$ and a $\bar{\vp}$. For instance, with $d=3$:
\begin{equation}\label{tinv}
\int d\vec{g} \phi(g_1,g_2,g_3)\bar{\phi}(g_1,g_2,g_3)=\int d\vec{g} d\vec{g}\,^{\prime} \phi(g_1,g_2,g_3)\left[\prod_{i=1}^3 \delta(g_i-g_i^\prime)\right]\bar{\phi}(g_1^\prime,g_2^\prime,g_3^\prime)\equiv \vcenter{\hbox{\includegraphics[scale=0.9]{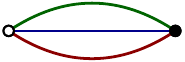} }}
\end{equation}
it is an example of  tensorial invariant made with two fields. We can see that any argument $g_i$ in $\vp$,  is summed with the corresponding argument on $\bar{\vp}$. The last term of the equation \eqref{tinv} introduce the diagramatic notation that we will use in the rest of this paper. Each interaction may be mapped into a regular bipartite colored graph, whose black (resp. white) vertices correspond to $\vp$ (resp. $\bar{\vp}$) fields, and colored line to Kronecker delta insertions between corresponding field variables. As maps over the circle, the field $\vp$ admits the Fourier transformation of the form
\bea
\vp(\vec g)=\sum_{\vec p\in \mZ^d} T_{\vec p} \exp[i(\vec p,\vec \theta)],\quad (\vec p, \vec \theta)=\sum_{j}p_j \theta_j,\quad  g_j=e^{i\theta_j}
\eea
where $T_{\vec p}$ stands for the Fourier mode. Then the action $S$  in the dual space $\mZ$ of $U(1)$ is written as:
\bea
S[T,\bar{T}]=\sum_{\vec{p}\in\mZ^d} \bar{T}_{\vec{p}}(\vec{p}\,^2+m^2)T_{\vec{p}} + \sum_{\vec{p}_j\in\mZ^d\,,\forall\,j} \mathcal{V}_{\vec{p}_1,\vec{p}_2,\vec{p}_3,\vec{p}_4} T_{\vec{p}_1}\bar{T}_{\vec{p}_2}T_{\vec{p}_3}\bar{T}_{\vec{p}_4}+\mathcal{O}(T^2,\bar{T}^2)\,,\label{action}
\eea
where $\vec{p}\,^2=\sum_i p_i^2$, and $\mathcal V$ is the interaction vertex or a sum of product of Kronecker deltas providing tensorial invariants, weighted with some coupling constants. As an example, for $d=4$:
\begin{equation}
\mathcal{V}^{(i)}_{\vec{p}_1,\vec{p}_2,\vec{p}_3,\vec{p}_4} := \lambda\,\delta_{p_{1i}p_{4i}}\delta_{p_{2i}p_{3i}}\prod_{j\neq i} \delta_{p_{1j}p_{2j}}\delta_{p_{3j}p{4j}}\equiv\lambda\,\vcenter{\hbox{\includegraphics[scale=0.7]{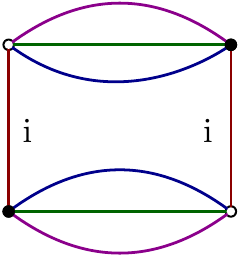} }}\,,\label{interaction4}
\end{equation}
is the \textit{melonic} interaction with intermediate lines of color $i$ (for a definition of melonic interactions and melonic diagrams, see \cite{Gurau:2009tw} and we will give a short definition in the next paragraph). The interactions are not local in the ordinary sens. However it has been showed that tensorial invariance provides an appropriate notion of locality for GFTs\footnote{In particular, this locality principle called traciality allows to define local counter-terms in the ordinary sens.} (see \cite{Carrozza:2013mna} for instance):
\begin{definition}
A connected tensorial invariant interaction is said to be local. In the same footing, any interacting action expanded as a sum of such diagrams is said to be local. 
\end{definition}
The statistical description of the model is given  by the partition function $Z(J,\bar{J})$:
\begin{equation}
Z(J,\bar{J})=\int dT d\bar{T} e^{-S[T,\bar{T}]+\bar{J}\cdot T+\bar{T}\cdot J}\,, \mbox{ with } \bar{A}\cdot B:= \sum_{\vec{p}} \bar{A}_{\vec{p}}B_{\vec{p}}\,.
\end{equation}
The Feynman rules allow to compute it in perturbation theory. In order to prevent UV divergences, the propagator $C(\vec{p}\,)=(\vec{p}\,^2+m^2)^{-1}$ has to be regularized. A current regularization scheme is the Schwinger regularization:
\bea
C^{-1}(\vec{p})=\int d\mu_{C_\Lambda}\,\,T_{\vec p}\,\bar T_{\vec p}=\frac{e^{-(\vec p\,^2+m^2)/\Lambda^2}}{\vec{p}\,^2+m^2},\,
\eea
Note that other regularization schemes holds. In the rest of this paper  we will consider the dimensional regularization, which consists as an extension of the dimension of the group manifold and exploit the analytic properties of the divergent amplitudes in term of this extension. \\

Let us conclude this section by providing some remarks about  the regularization procedure. We do not make any explicit choice for the regularization of the UV divergences, and many expressions which appears in this paper are superficially divergent. To be more precise, we have to think that all the divergences are regularized from a specific dimensional regularization procedure and  such that any  expressions are analytically continued in the dimension of the group manifold. In practice, this corresponds to the replacement   $U(1)\to U(1)^D$ as an analytically continuation on the parameter $D$ and have been considered in \cite{Carrozza:2013wda}-\cite{Geloun:2013saa}. Using such regularization procedure leads to a nice simplification in the computation of the WT-identities and exhibit the term of the form  $C(\vec{p}\,)-C(\vec{p}\,^{\prime})$, which could be complicated to trait  using other regularization schemes.

\subsection{Renormalizability and renormalization group flow}
The model defined with the action \eqref{action} is showed  to be just-renormalizable for $d=5$ \cite{Carrozza:2012uv} and the interaction restricted to quartic melonic interactions. The model is defined with the action of the form
\begin{equation}
S[T,\bar{T}]=\sum_{\vec{p}\in\mZ^5} \bar{T}_{\vec{p}}(\vec{p}\,^2+m^2)T_{\vec{p}} + \lambda\sum_{i=1}^{d=5}\sum_{\vec{p}_j\in\mZ^5\,,\forall\,j} \mathcal{W}^{(i)}_{\vec{p}_1,\vec{p}_2,\vec{p}_3,\vec{p}_4} T_{\vec{p}_1}\bar{T}_{\vec{p}_2}T_{\vec{p}_3}\bar{T}_{\vec{p}_4}\,,
\end{equation}
where $\mathcal{W}^{(i)}$ corresponds to the diagram of equation \ref{interaction4}, and all the divergences occurring on the amplitudes may be removed at all order with a finite number (three) of counter-terms: one for mass, one for coupling, and the last one for wave function; $Z_m$, $Z_\lambda$ and $Z$ respectively. The \textit{renormalized parameters} $\lambda_r$ and $m_r^2$ are then defined as:
\begin{equation}
(T,\bar{T})\to Z^{1/2}(T,\bar{T})\,,\quad m^2=Z^{-1}Z_mm_r^2 \,,\quad \lambda=Z^{-2}Z_\lambda\lambda_r\,,
\end{equation}
such that all the amplitudes computed from the partition function built in term of the renormalized quantities are finite order by order in the perturbative expansion. In the classical point of view  the renormalized  action becomes:
\begin{equation}
S[T,\bar{T}]=\sum_{\vec{p}\in\mZ^5} \bar{T}_{\vec{p}}(Z\vec{p}\,^2+Z_mm^2_r)T_{\vec{p}} +Z_\lambda \lambda_r\sum_{i=1}^{d=5}\sum_{\vec{p}_j\in\mZ^5\,,\forall\,j} \mathcal{W}^{(i)}_{\vec{p}_1,\vec{p}_2,\vec{p}_3,\vec{p}_4} T_{\vec{p}_1}\bar{T}_{\vec{p}_2}T_{\vec{p}_3}\bar{T}_{\vec{p}_4}\,.
\end{equation}\label{Sclassic}
The divergent parts of the counter-terms are fixed from the requirement that they cancels the UV-loop-divergences. The finite parts however are fixed from renormalized conditions: see   \eqref{propcounterterms}. We will  trait in detail  the choice of these renormalization conditions in the next sections. \\

\noindent
The renormalization group flow describes the change of the couplings in the effective action when UV degrees of freedom are integrated out. The functional renormalization group formalism  is a specific way to build such an evolution in the parameter space. To make the partition function dynamic, let us consider the following  parameter ($s\in \mathbb{R}$) such that $Z(J,\bar{J})$ is replaced by $Z_s(J,\bar{J})$:
\begin{equation}
Z_s(J,\bar{J})=\int dT d\bar{T} e^{-S[T,\bar{T}]-R_{s}[T,\bar{T}]+\bar{J}\cdot T+\bar{T}\cdot J}\,,
\end{equation}
where for  the rest we denote by $S_{source}(T,\bar{T}, J,\bar{J}):=\bar{J}\cdot T+\bar{T}\cdot J$, and where the regulator $R_{s}[T,\bar{T}]$ is chosen of the form:
\begin{equation}
R_{s}[T,\bar{T}]:=\sum_{\vec{p}} r_{s}(\vec{p}\,^2) \bar{T}_{\vec{p}}T_{\vec{p}}\,. 
\end{equation}
Note that the regulator may be included in a suitable redefinition of the bare propagator:
\begin{equation}
C^{-1}(\vec{p})\to C^{-1}_{s}(\vec{p})=\frac{1}{Z\vec{p}\,^2+r_{s}(\vec{p}\,^2)+Z_mm^2_R}\,.
\end{equation}
The function  $r_{s}$ satisfy the standard properties of the Wetterich-Morris regulator, that we recall here for convenience for the reader: 
\begin{itemize}
\item $r_{s}\geq 0$,
\item $\lim_{s\rightarrow -\infty}r_{s}=0$,
\item $\lim_{s\rightarrow \infty}r_{s}=\infty\,.$
\end{itemize}
To be more precise,  $\Lambda\gg1$ is an UV-cutoff, and corresponds to the microscopic scale at which the classical action is well defined and playing the role of an initial condition. When $s$ goes from $\ln(\Lambda)$ to $-\infty$, the effective cutoff $e^s$ run from $\Lambda$ (UV regime) to $0$ (IR regime).

The central object of the FRG approach is the effective action, which is defined as the Legendre transform of the standard free energy. The free energy itself, $W_s$ is defined as: 
\begin{equation}
W_s:=\ln(Z_s).
\end{equation}
The regulator modify the definition of the effective action used in the FRG formalism, called \textit{average effective action} and denoted by $\Gamma_s$. It is a functional depending on the means fields $M$ and $\bar{M}$ 
\begin{equation}
M_{\vec{p}}:=  \frac{\partial W}{\partial \bar{J}_{\vec{p}}}\,, \quad \bar{M}_{\vec{p}}:=  \frac{\partial W}{\partial {J}_{\vec{p}}}\,,
\end{equation}
and defined as
\begin{equation}
\Gamma_{s}[M,\bar{M}]+R_{s}[M,\bar{M}]=\bar{J}\cdot M+\bar{M}\cdot J-W_{s}[J,\bar{J}]\,.
\end{equation}
This transformation is said to be  a modified Legendre transformation because of the presence of the regulator on the left hand side. This definition ensures that the effective average action satisfies the boundary conditions
\begin{equation}
\Gamma_{s=\ln(\Lambda)}=S\,,\qquad \Gamma_{s=-\infty}=\Gamma\,,
\end{equation}
where $\Gamma$ is  the standard effective action, i.e. the  Legendre transform of the free energy without regulator. In contrast with ordinary presentations, we choose the initial conditions  $S$ so that the perturbative expansion built from $\Gamma$ (or equivalently from the original partition function) is free of divergences. Obviously, all loop divergences for arbitrary $s$ are canceled by the counter-terms computed for $e^s=0$ (because the regulator behaves has an effective mass, which decreases the weight of each propagator lines). As a result, it is suitable to fix the renormalization conditions for $s=-\infty$, so that all the counter-terms do not depend on $s$. The difference between loops at arbitrary $s$ and counter-terms provide from the  effective mass, coupling and wave-function and  whose evolution is governed by the so called \textit{Wetterich equation}. We will define and scrutinize very clearly the effective parameter in Section \ref{section4} and we will give  the motivations for these initial conditions in the next section.\\

\noindent
The evolution of the effective average action $\Gamma_{s}$ with the renormalization scale $s$ is governed by an exact   flow  given by the Wetterich equation \cite{Wetterich:1989xg}:
\begin{equation}\label{wetterich}
\partial_s \Gamma_{s}=\sum_{\vec{p}}\partial_s r_{s}(\vec{p}\,)(\Gamma^{(2)}_{s}+r_{s})^{-1}(\vec{p},\vec{p}\,)\,,
\end{equation}
where $\Gamma^{(2)}_{s}$ is the second functional derivative
respect to the mean fields or inverse propagator in the presence of arbitrary fields $\Gamma^{(2)}_{s}:= \frac{\partial^2\Gamma_s}{\partial M\partial\bar{M}}$. Note that the inverse of the two-point function (the effective propagator) is $\Gamma^{(2)}_{s}+r_{s}$. The infrared cutoff $r_{s}$ should guarantee that only a small momentum range  contributes to \eqref{wetterich}, and such  that the r.h.s. is ultraviolet and infrared finite. Despite its simple structure, Equation \eqref{wetterich} is a complicated non-linear functional differential equation, which generally requires some approximation in order to be solved. To close this section, we make a little change in the notation. At this stage and for the rest of the paper, we will denote by $Z_{-\infty}$ the wave function counter-term, that we called $Z$ in the previous section. The ``$-\infty$'' recalling that finite part is fixed in the infrared limit. We will talk about this point    in Section \ref{section4}. \\

\noindent
To close this section, we recall for the reader a central notion for our purpose the notion of \textit{canonical dimension}. In standard quantum field theories, interactions are classified following their proper dimension (coming from the definition of the action over a $d$-dimensional differential manifold). Interactions with positive or zero dimensions are said to be \textit{renormalizable}, while interactions with negative dimensions are said to be \textit{non-renormalizable}. For TGFTs, such classification holds, but the notion of dimension is more subtle. Indeed, strictly speaking, there are no dimensions in the action \eqref{Sclassic} (the sums over $\mathbb{Z}^d$ are dimensionless in contract with integration over space-time in standard quantum field theory). The notion of canonical dimension emerge from the perturbative expansion, and basically corresponds to the way in which the radiative corrections occurs. In some words, one can show that all the radiative corrections in the leading sector of the perturbative expansion for the $4$-point function are logarithmically divergent. Then, it is coherent to associate an exponent zero for the scaling of the effective couplings, and this is the scaling that we call canonical dimension. The dimension of the quartic coupling being fixed, we can consider the loop corrections for the mass, and we conclude that this parameter scale with a power $2$ of the cut-off. Then, we fix the dimensions of $m^2$ and $\lambda$ as:
\begin{equation}
[\lambda]=0\,,\qquad [m^2]=2\,.
\end{equation}
For more informations about canonical dimension, the reader may consult \cite{Benedetti:2014qsa}.

\section{Symmetric and non-symmetric phases}\label{section3}
We start this section with a definition and a discussion about a crucial notation  for our results:
\begin{definition}
\textbf{Symmetric and non-symmetric phases in the melonic sector} - As long as the effective two-points function $G_s$ remains diagonal: $G_{s,\vec{p}\vec{q}}=G_{s}(\vec{p})\delta_{\vec{p}\vec{q}}$, the theory is said to be in the symmetric or perturbative phase.  If this is not the case the theory is said to be in the non-symmetric or non-perturbative regime.
\end{definition}

\noindent
Note that this definition make sens. First, the equivalences between \textit{symmetric} and \textit{perturbative} on one hand, and \textit{non-symmetric} and \textit{non-perturbative} on the other hand can be deduced easly. Indeed, in the perturbation theory, the effective propagator is obtained from the computation of the \textit{leading-order} self-energy $\Sigma$ in the UV as a sum of Feynman graphs, as follow:\\
\begin{equation}
\Sigma=\left\{\vcenter{\hbox{\includegraphics[scale=0.6]{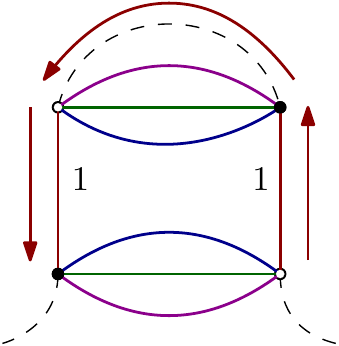} }} +\cdots \right\}+\left\{\vcenter{\hbox{\includegraphics[scale=0.6]{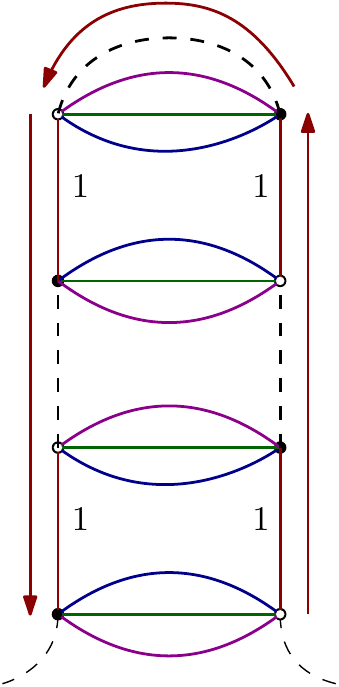} }} +\vcenter{\hbox{\includegraphics[scale=0.6]{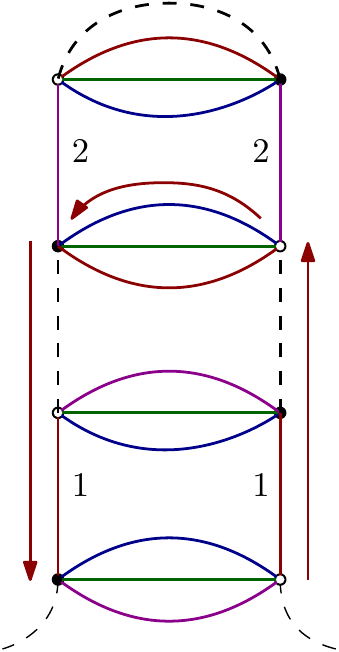} }}\cdots \right\}+\mathcal{O}(\lambda^3)\,,\label{eqexp}
\end{equation}
\noindent
where the dot lines correspond to the Wick contractions with the propagator $C_s$ according to the Gaussian measure, and the half dot lines correspond to the external lines of the Feynman graphs. Note that such a Feynman graph, strictly speaking corresponds to a $2$-complexes rather than ordinary Feynman graphs, i.e.  gluing sets of vertices, lines and faces. The vertices and lines are the sets of black and white nodes and the set of colored lines, including lines of color 0 conventionally attributed to dashed lines. We recall the following definition:\\
\begin{definition}
A face is defined as a maximal and bicolored connected subset of lines, necessarily including the color 0. We have two kind of faces:\\

\noindent
$\bullet$ The closed or internal faces, when the bicolored connected set correspond to a cycle.\\

\noindent
$\bullet$ The open or external faces when the bicolored connected set does not close as a cycle.\\
\end{definition}
Moreover, we recall that the leading order graphs have a well known recursive structure and are well known as \textit{melonic diagrams} \cite{Gurau:2010ba}. Now, for each diagrams, the melonic configuration identify $d-1$ external momenta over $d$ at the external vertex, and the last ones from the momentum conservation along the boundary of the external faces (pictured as the red arrow path). Then, the perturbative melonic $2$-point function is necessarily a diagonal function. Furthermore, non-perturbative effects may be a source of deviation from this structure. To see why, we can simply consider the truncation around quartic interaction. In the perturbative theory, the means field vanish. However a non-vanishing means field provides from a non-perturbative effect. But if the means field does not vanish, non-symmetric terms occurs simply dues to the contribution of the quartic interaction term in the average effective action.
\begin{corollary}
The symmetric phase corresponds to a vanishing means field, whereas the non-symmetric phase corresponds to non-vanishing means field.
\end{corollary}
\begin{corollary}
In the symmetric phase regime, all the odd-functions vanish.
\end{corollary}
 We will recall some properties of the leading order diagrams, called \textit{melons} (the reader may be consult \cite{Gurau:2010ba} and \cite{Geloun:2013saa} for more details). We recall that for any Feynman graph $\mathcal{G}$, the perturbative power counting is given by:
\begin{equation}
\omega(\mathcal{G})=-2L(\mathcal{G})+F(\mathcal{G})\,,
\end{equation}
where $L$ and $F$ denote the sets of and internal faces. 
\begin{definition} \textbf{(Melonic diagrams)}\,
At fixed number of external lines, said $2N$ ($N\in\mathbb{N}$), the graphs for which the power counting is optimal are said melonics. 
\end{definition}
\begin{proposition} \textbf{(Recursive construction of vacuum melons)}\, 
For the quartic melonic model, the elementary vacuum melons denoted $g_i$ and built of a single vertex are the following\\
\begin{equation}
g_i=\vcenter{\hbox{\includegraphics[scale=0.6]{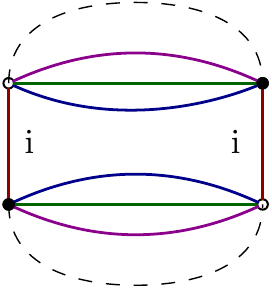} }} \,.
\end{equation}
Then, higher melons are obtained recursively from elementary melon by the replacements:
\begin{center}
\includegraphics[scale=0.6]{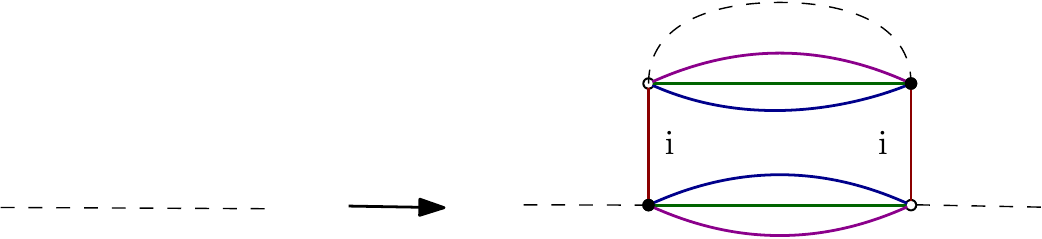} 
\end{center}
for arbitrary $i$.
\end{proposition}
The figure \ref{figmelons} provide an illustration of this explicit construction. The proposition can be easily proved by induction. For more details the reader may be consult standard references as \cite{Carrozza:2014rba}. \\
\begin{center}
\includegraphics[scale=0.6]{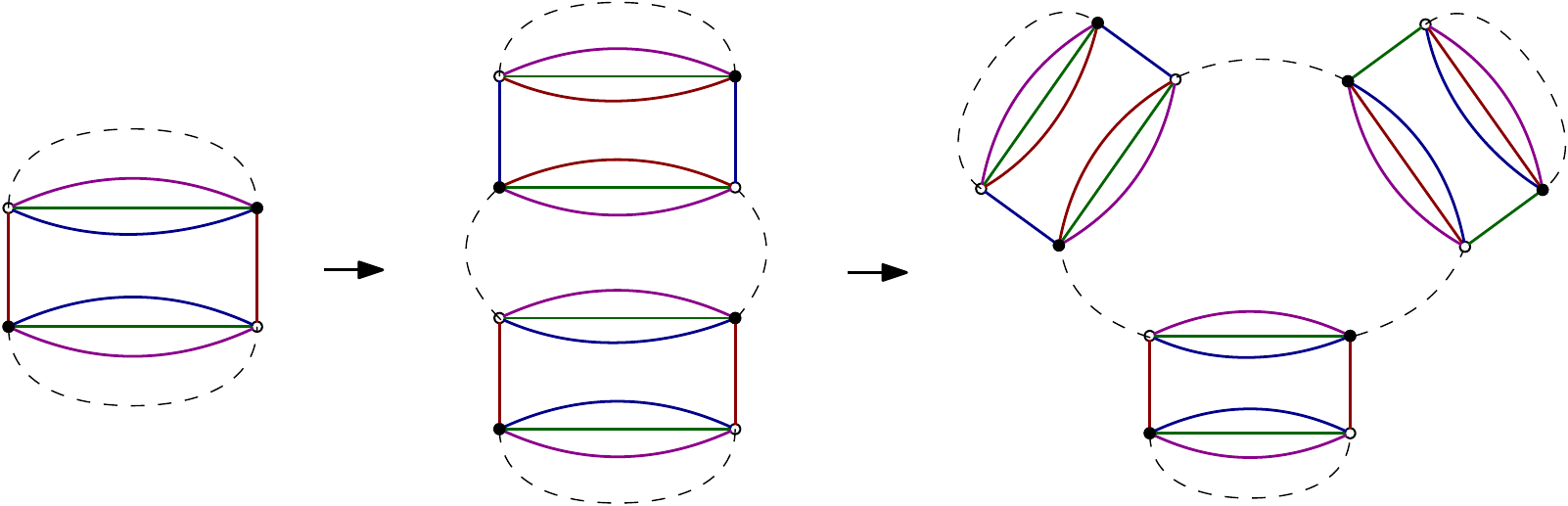} 
\captionof{figure}{Example of melonic vacuum diagrams built from an elementary vacuum diagram. It is easy to cheek that all these diagrams have the same power counting.}
\end{center}\label{figmelons}

\noindent
1PI-non-vacuum vertices may be obtained from vacuum vertices cutting some dotted lines in such a way that the divergent degree of the resulting graph remains optimal. For a fixed number of external lines, the \textit{1PI-non-vacuum melons} are then defined as the graphs with highest divergent degree in this procedure. Figure \ref{fig1} provides an example for $6$-points functions. 
\begin{center}
\includegraphics[scale=0.6]{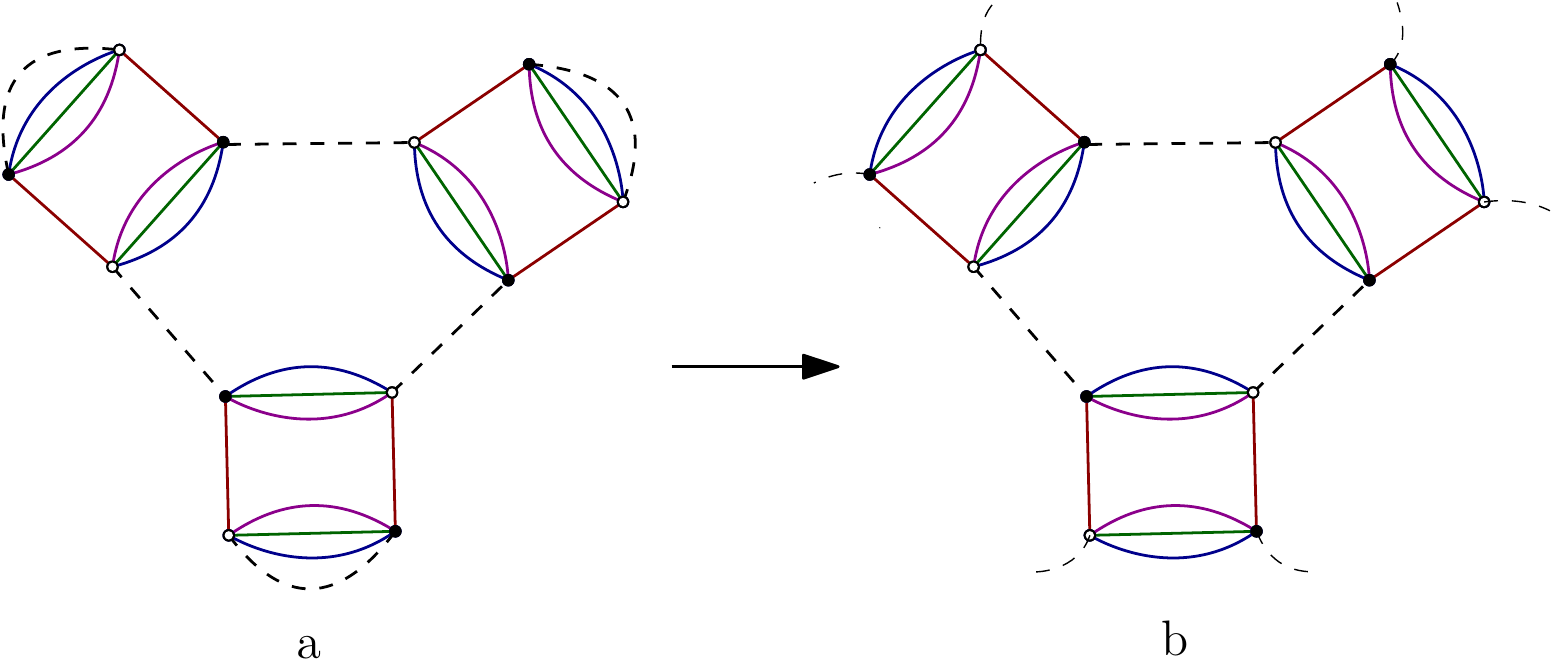} 
\captionof{figure}{A melonic vacuum graph (a) and a leading order $6$-point function (b). Cutting the three internal lines discard $3(d-1)$ faces of length $1$, and a red face of length $3$ having the three discarded lines as boundary. }\label{fig1}
\end{center}
Starting from the vacuum melon on figure \ref{fig1}a, we have to cut three lines to obtain a graph with $6$ external lines. It is easy to see that we can not cut a line of the central loop of length three. Indeed, if we cut such a line, we obtain a 1PR graph. Then, the only way is to cut tadpole lines. The first cutting delete $d$ faces, while 
the two second deletes $2(d-1)$ faces. The reason is that the first cutting has deleted $d-1$ faces of length one, and a long face of length three passing through all the vertices, which becomes three external lines of the same color passing through the internal lines. Note that it is possible only because all the vertices are of the same type. The procedure may be easily generalized. To this end, we start with some definitions:

\begin{definition} \textbf{Boundary and heart vertices and external lines}\\

\noindent
$\bullet$ Any vertex hooked with an external line is said to be a boundary vertex. Other vertices are said heart vertices\\

\noindent
$\bullet$ Any external faces passing through a single external vertex is said to be an boundary external faces. \\

\noindent
$\bullet$ Any external faces passing through at least one heart vertex is said to be an heart external faces.
\end{definition}

\begin{definition}
The heart graph of a melonic 1PI Feynman graph $\mathcal{G}$ is the sub-graph (i.e. the subset of vertices and lines) obtained from deletion of the external vertices. 
\end{definition}

\noindent
Now, consider a vacuum melonic diagram. We obtain a $2$-point graph cutting one of the dotted lines. Because of the structure of melonic diagrams, it is clear that if we cut a line which is not a tadpole line (i.e. a line in a loop of length upper than one), we obtain a 1PR diagrams. Then, we have to cut only tadpole lines. Cutting the first one, we delete $d$ faces, $d-1$ become boundary external lines while the other one becomes an heart external lines. We have then obtain a 1PI $2$-points melonic diagram. To obtain a $4$-points melonic diagram, we have to cut another tadpole line on this diagram. However, it is clear that such a cutting could be deleted $d$ internal faces, \textit{except} if the chosen vertex share the opened heart external face. Indeed, in this case, the cutting cost $d-1$ faces (which become boundary external lines) for the same cost in dotted lines, and the power counting is clearly optimal. Moreover, the heart external face of the original $2$-point diagram become two heart external faces, clearly of the same colors. Recursively, we deduce the following proposition:
\noindent
\begin{proposition} \label{cormelons}
A 1PI melonic diagram with $2N$ external lines has $N(d-1)$ external faces of length $1$ shared by external vertices and $N$ heart external faces of the same color running through the internal vertices and/or internal lines (i.e. through the heart graph). 
\end{proposition}
The figure \ref{figmelons} and the graphs on equation \eqref{eqexp} give some examples. On figure \ref{figmelons} three red external faces run through internal lines, and on the graphs of equation \eqref{eqexp}, the red external face run through internal lines and vertices. To complete these definitions, and of interest for our incoming results, we have the following proposition:
\begin{proposition}
For  all the model that we consider, all the divergences are contained in the melonic sector. 
\end{proposition}
Let us point out that  all the counter-terms in the perturbative renormalization are fixed from the melonic diagrams only. A proof may be found in \cite{BenGeloun:2011rc}. Finally, we can  add an important remark about melonic diagrams: Their divergent degrees depends only on their external lines. This is expected for a renormalizable theory. To be more precise, note that the number of line is related to the number of vertices as $2L=4V-N_{ext}$, where $N_{ext}$ denotes the number of external lines. Moreover, it is easy to see, from the recursive definition of melons that $F=4(L-V+1)$ \cite{Samary:2012bw}. Indeed, contracting a tree line does not change the divergent degree and the number of faces. Then, contracting all the line over a spanning tree, we get $L-V+1$ lines contracted over a single vertex. Now, we delete the lines optimally. We have some external lines, but we know from the definition of melons that no more one 
heart external face pass through one of them. Then, an optimal cutting is for a line which is on the boundary of one external face. As a result, the cutting remove $4$ internal lines. By processing the operation until the last line have been contracted, we find the counting for faces and therefore the divergence degree becomes
\begin{equation}\label{melocountinplus}
\omega=-2L+F=-2(2V-N_{ext}/2)+4(V-N_{ext}/2+1)=4-N_{ext}\,.
\end{equation}

\section{Ward-Takahashi identities}\label{section4}
Symmetries in quantum field theories give rise the relations between various Green's functions and, therefore, between the transition amplitudes. For TGFT, WT-identities come from the unitary invariance of the interactions in the classical theory. It is expected that the WT-identities may introduce non-trivial constraint on the renormalization group flow, that we could take into account for an improvement of standard non-perturbative methods in the TGFT context. In fact, we will see in the next section that, in the leading order sector in the deep UV limit, Ward identities share exactly the same information that \textit{melonic structure equations} discussed in the next section.  \\ 

\noindent
Let $\mathcal{U}=(U_1,U_2,\cdots, U_d)$, where the $U_i\in \textit{Unit}_\infty$ are infinite size unitary matrices in momentum representation. We define the transformation (we use the Einstein convention for indices summation):
\bea
\mathcal{U}[T]_{p_1,p_2,\cdots, p_d}&=& \sum_{p_1',\cdots,p_d'}U_{1\,,p_1p_1^\prime}U_{2\,,p_2p_2^\prime}\cdots U_{d\,,p_dp_d^\prime}                                           T_{p_1^\prime,p_2^\prime,\cdots, p_d^\prime}\cr
&=:&U_{1\,,p_1p_1^\prime}U_{2\,,p_2p_2^\prime}\cdots U_{d\,,p_dp_d^\prime}                                           T_{p_1^\prime,p_2^\prime,\cdots, p_d^\prime}\,,\label{transform}
\eea
such that the interaction term is invariant i.e.
\begin{equation}
\mathcal{U}[S_{int}]=S_{int}\,.
\end{equation}
However the kinetic term is not left invariant with $\mathcal{U}$. The formal invariance of the  path integral then imply that the variations of these terms have to be compensate by a non trivial variation of the source terms. Then, we will investigate the effect of an infinitesimal variation, and let us consider an infinitesimal transformation: $\mathcal{U}=\mathbb{\textbf{I}}+\vec{\epsilon}$, with :
\begin{equation}
\vec{\epsilon}=\sum_i\mathbb{I}^{\otimes (i-1)}\otimes \epsilon_i\otimes \mathbb{I}^{\otimes(d-i)}\,,
\end{equation}
where $\mathbb{I}$ is the identity on $\textit{Unit}_\infty$, $\mathbb{\textbf{I}}=\mathbb{I}^{\otimes d}$ the identity on $\textit{Unit}_\infty^{\,\otimes d}$, and $\epsilon_i$ denotes skew-symmetric hermitian matrix such that $\epsilon_i=-\epsilon_i^\dagger$ and 
\begin{equation}
\vec{\epsilon}_i[T]_{\vec{p}}={\epsilon_i}_{p_ip_i^\prime}T_{p_1,\cdots,p_i^\prime,\cdots, p_d}\,.
\end{equation}
The  invariance of the path integral means $\vec{\epsilon}\,[Z_s[J,\bar{J}]]=0$, i.e.:
\begin{equation}\label{sat1}
\vec{\epsilon}\,[Z_s[J,\bar{J}]]=\int dT d\bar{T} \bigg[ \vec{\epsilon}\,[S_{kin}]+\vec{\epsilon}\,[S_{int}]+\vec{\epsilon}\,[S_{source}]\bigg] e^{-S_s[T,\bar{T}]+\bar{J}T+\bar{T}J} =0.\,
\end{equation}
Computing each term separately, we get successively using linearity of the operator $\vec{\epsilon}$:
\begin{equation}
\vec{\epsilon}\,[S_{int}]=0\,,
\end{equation}
\bea
\vec{\epsilon}\,[S_{source}]=-\sum_{i=1}^d\sum_{\vec{p}, \vec{p}\,^{\prime}} \prod_{j\neq i} \delta_{p_jp_j^\prime} [\bar{J} _{\vec{p}}\,T_{\vec{p}\,^\prime}-\bar{T}_{\vec{p}}{J} _{\vec{p}\,^\prime}]{\epsilon_i}_{p_ip_i^\prime}\,,
\eea
\bea
 \vec{\epsilon}\,[S_{kin}]=\sum_{i=1}^d\sum_{\vec{p}, \vec{p}\,^{\prime}} \prod_{j\neq i} \delta_{p_jp_j^\prime} \bar{T}_{\vec{p}}\big[C_s(\vec{p}\,^{2})-C_s(\vec{p}\,^{\prime\,{2}})\big]T_{\vec{p}\,^\prime}{\epsilon_i}_{p_ip_i^\prime}\,,\label{sat2}
\eea
 Combining the two expressions \eqref{sat1} and \eqref{sat2}, we come to 
\begin{align}
\sum_{i=1}^d\sum_{\vec{p}, \vec{p}\,^{\prime}} \prod_{j\neq i}  \delta_{p_jp_j^\prime}  \bigg[\frac{\partial}{\partial J_{\vec{p}} }\big[C_s(\vec{p}\,^{2})-C_s(\vec{p}\,^{\prime\,{2}})\big]\frac{\partial}{\partial \bar{J}_{\vec{p}\,^\prime}}-\bar{J} _{\vec{p}}\,\frac{\partial}{\partial \bar{J}_{\vec{p}\,^\prime}}+{J} _{\vec{p}\,^\prime}\frac{\partial}{\partial J_{\vec{p}} }\bigg] e^{W_s[J,\bar{J}]} =0\,,\label{socle2}
\end{align}
where we used  the fact that, for all polynomial $P(T,\bar T)$ the following identity holds:
\bea
\int\, d\mu_C\,\, P(T,\bar T) e^{\bar J T+\bar T J}=\int\, d\mu_C\,\, P(\frac{\partial}{\partial \bar J},\frac{\partial}{\partial J}) e^{\bar J T+\bar T J}.
\eea
 Equation \eqref{socle2} is satisfied for all $i$. Then, expanding each derivative, and setting $i=1$, we deduce the following theorem:\\

\begin{theorem}
The partition function $Z_s[J,\bar{J}]=:e^{W_s[J,\bar{J}]}$ of the theory defined by the action \ref{Sclassic} verify the following  (WT identity),
\bea\label{left}
\sum_{\vec{p}_\bot, \vec{p}_\bot\,^{\prime}} \prod_{j\neq 1}  \delta_{p_jp_j^\prime}  \bigg\{\big[C_s(\vec{p}\,^{2})-C_s(\vec{p}\,^{\prime\,{2}})\big]\left[\frac{\partial^2 W_s}{\partial \bar{J}_{\vec{p}\,^\prime}\,\partial {J}_{\vec{p}}}+\bar{M}_{\vec{p}}M_{\vec{p}\,^\prime}\right]-\bar{J} _{\vec{p}}\,M_{\vec{p}\,^\prime}+{J} _{\vec{p}\,^\prime}\bar{M}_{\vec{p}}\bigg\}=0\,,\label{Ward0}
\eea
with $\vec{p}_\bot:=(0,p_2,\cdots ,p_d)\in \mathbb{Z}^{d}$.
\end{theorem}
WI-identity contains some informations on the relations between Green functions. In particular, they provide a relation between $4$ and $2$ points functions, which,  maybe translated as a relation between wave function renormalization $Z$ and vertex renormalization $Z_\lambda$. Applying $\partial^2/\partial M_{\vec{q}\,^\prime}\,\partial \bar{M}_{\vec{q}}$ on the left hand side of \eqref{left}, and taking into account the relations
\begin{equation}
\frac{\partial M_{\vec{p}}}{\partial {J}_{\vec{p}\,^{\prime}}}= \frac{\partial^2 W_s}{\partial \bar{J}_{\vec{p}}\,\partial {J}_{\vec{p}\,^{\prime}}}\, \quad\mbox{and }\quad  \, \frac{\partial \Gamma_s}{\partial M_{\vec{p}}}=\bar{J}_{\vec{p}}-r_s(\vec{p})\bar{M}_{\vec{p}}\,,
\end{equation}
as well as the definition $G_{s\,,\vec{p}\vec{p}\,^\prime}^{-1}:=(\Gamma^{(2)}_s+r_s\big)_{\vec{p} \vec{p}\,^\prime}$, we find that
\begin{align}
\nonumber\sum_{\vec{p}_\bot, \vec{p}_\bot\,^{\prime}} \prod_{j\neq 1}  \delta_{p_jp_j^\prime}  &\bigg[\big[C_s(\vec{p}\,^{2})-C_s(\vec{p}\,^{\prime\,{2}})\big]\bigg[\frac{\partial^2 G_{s\,,\vec{p},\vec{p}\,^\prime}}{\partial M_{\vec{q}\,^\prime}\,\partial \bar{M}_{\vec{q}}}+\delta_{\vec{p}\vec{q}}\,\delta_{\vec{p}\,^\prime,\vec{q}\,^\prime}\bigg]-\Gamma^{(2)}_{s\,,\vec{q}\vec{p}}\,\delta_{\vec{q}\,^\prime\vec{p}\,^\prime}+\Gamma^{(2)}_{s\,,\vec{q}\,^\prime\vec{p}\,^\prime}\delta_{\vec{p}\vec{q}}\\
&-r_s(\vec{p}\,^2)\delta_{\vec{q}\vec{p}}\,\delta_{\vec{q}\,^\prime\vec{p}\,^\prime}+r_s(\vec{p}^{\,\prime\,2})\delta_{\vec{q}\,^\prime\vec{p}\,^\prime}\delta_{\vec{p}\vec{q}}-\Gamma_{s,\vec{q};\vec{q}^\prime\vec{p}}^{(1,2)}\,M_{\vec{p}\,^\prime}+\Gamma^{(2,1)}_{s,\vec{q}\vec{p}^\prime;\vec{q}^\prime}\bar{M}_{\vec{p}}\bigg] =0\,,
\end{align}
where we have introduced the following notation:
\begin{equation}
\Gamma^{(n,m)}_{s,\vec{p}_1,\cdots,\vec{p}_n;\vec{p}_1\cdots\vec{p}_m}=:\frac{\partial^{m+n}\Gamma_s}{\partial M_{\vec{p}_1}\cdots\partial M_{\vec{p}_n}\partial\bar{M}_{\vec{{p}}_1}\cdots\partial\bar{M}_{\vec{{p}}_m}},\, \Gamma^{(2)}_s:=\Gamma_s^{(1,1)}.
\end{equation}
Now setting
\begin{equation}
\frac{\partial^2 G_{s\,,\vec{p},\vec{p}\,^\prime}}{\partial M_{\vec{q}\,^\prime}\,\partial \bar{M}_{\vec{q}}}=-G_{s\,,\vec{p}\vec{n}}\tilde{\Gamma}^{(4)}_{s,\vec{n},\vec{q};\vec{m},\vec{q}\,^\prime}G_{s\,,\vec{m}\vec{p}\,^\prime}\,,
\end{equation}
where we use Einstein summation for repeated indices.  We come to
\begin{equation}
\tilde{\Gamma}^{(4)}_{s,\vec{n},\vec{m};\vec{p},\vec{q}}:={\Gamma}^{(4)}_{s,\vec{n},\vec{m};\vec{p},\vec{q}}\,-\Gamma^{(2,1)}_{s,\vec{n}\vec{m};\vec{r}}G_{s,\vec{r}\vec{s}}\Gamma_{s,\vec{s};\vec{p}\vec{q}}^{(1,2)}-\Gamma_{s,\vec{m};\vec{p}\vec{r}}^{(1,2)}G_{s,\vec{r}\vec{s}}\Gamma^{(2,1)}_{s,\vec{s}\vec{n};\vec{q}}\,.
\end{equation}
Finally, we deduce the following relation that we called \textit{First Ward-Takahashi identity}:
\begin{corollary} \textbf{First Ward-Takahashi identity}
\begin{align}
\nonumber\sum_{\vec{p}_\bot, \vec{p}_\bot\,^{\prime}} \prod_{j\neq 1}  \delta_{p_jp_j^\prime} & \bigg[\Delta C_s(\vec p,\vec p\,')\bigg[G_{s\,,\vec{p}\vec{n}}\tilde{\Gamma}^{(4)}_{s,\vec{n},\vec{q};\vec{m},\vec{q}\,^\prime}G_{s\,,\vec{m}\vec{p}\,^\prime}-\delta_{\vec{p}\vec{q}}\,\delta_{\vec{p}\,^\prime\vec{q}\,^\prime}\bigg]-\Gamma^{(2)} _{s\,,\vec{q}\vec{p}}\,\delta_{\vec{q}\,^\prime\vec{p}\,^\prime}+\Gamma^{(2)}_{s\,,\vec{q}\,^\prime\vec{p}\,^\prime}\delta_{\vec{p}\vec{q}}\\&-r_s(\vec{p}\,^2)\delta_{\vec{q}\vec{p}}\,\delta_{\vec{q}\,^\prime\vec{p}\,^\prime}+r_s(\vec{p}^{\,\prime\,2})\delta_{\vec{q}\,^\prime\vec{p}\,^\prime}\delta_{\vec{p}\vec{q}}-\Gamma_{s,\vec{q};\vec{q}\,^\prime\vec{p}}^{(1,2)}\,M_{\vec{p}\,^\prime}+\Gamma^{(2,1)}_{s,\vec{q}\vec{p}\,^\prime;\vec{q}\,^\prime}\bar{M}_{\vec{p}}\bigg] =0\,.\label{Ward}
\end{align}
with:
\begin{equation}
\Delta C_s(\vec p,\vec p\,'):=C_s(\vec{p}\,^{\prime\,{2}})-C_s(\vec{p}\,^{\,{2}})\,.
\end{equation}
\end{corollary}

\noindent
For the rest of this paper, we will consider  only  the WT-identity in the symmetric phase, which is given  for vanishing mean field. In this sector, the $3$-point functions vanish. Moreover, in the melonic sector, the $4$-point function has the following structure:
\begin{equation}
\Gamma_{s,\vec{p}\vec{q};\vec{r}\vec{s}}^{(4)}=\sum_{i=1}^d\vcenter{\hbox{\includegraphics[scale=0.7]{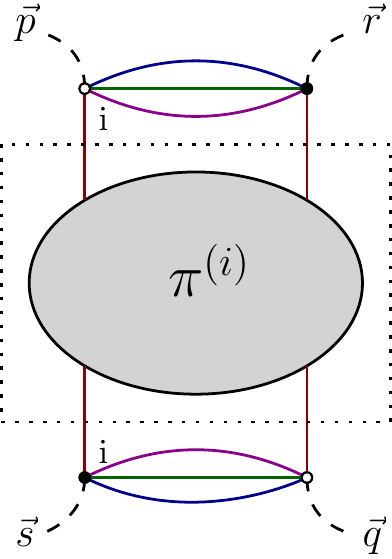}  }}+\vec{p}\leftrightarrow\vec{q}\equiv\sum_{i=1}^d \pi^{(i)}_{p_iq_i}\sym \mathcal{W}^{(i)}_{\vec{p},\vec{r},\vec{q},\vec{s}}=:\sum_i\Gamma_{s,\vec{p}\vec{q};\vec{r}\vec{s}}^{(4)\,,i}\,,\label{eqfour}
\end{equation}
with:
\begin{equation}
\sym \mathcal{W}^{(i)}_{\vec{p},\vec{r},\vec{q},\vec{s}}:= \mathcal{W}^{(i)}_{\vec{p},\vec{r},\vec{q},\vec{s}}+ \mathcal{W}^{(i)}_{\vec{q},\vec{r},\vec{p},\vec{s}}\,.
\end{equation}
This structure come from the definition of melonic diagrams. Let us recall that in Corollary \ref{cormelons} the melonic diagrams with external lines necessarily have internal faces of the same color running in the interior of the diagrams, while other external faces remain at the level of external vertices. In \eqref{eqfour}, these internal lines have a  color i (in red). Note that in \eqref{eqfour}, $\pi^{(i)}_{p_iq_i}$ depends \textit{à priori} on the four external variables, that is $p_i, q_i, r_i, s_i$, but the momentum conservation along the boundaries of the external faces ensure that $p_i=s_i$ and $q_i=r_i$. This quantity is  called \textit{effective vertex function}. Now, due to the fact that in the symmetric phase: $G_{s\,,\vec{q}\vec{p}}=G_{s}(\vec{p})\delta_{\vec{q}\,\vec{p}\,}$, the left hand side of the Ward identity \ref{Ward} take the form:
\begin{equation}
\sum_{\vec{p}_\bot,\vec{p}^\prime_\bot} \delta_{\vec{p}_\bot,\vec{p}^\prime_\bot}(C_s(\vec{p}\,)-C_s(\vec{p}\,^\prime))\left\{\vcenter{\hbox{\includegraphics[scale=0.5]{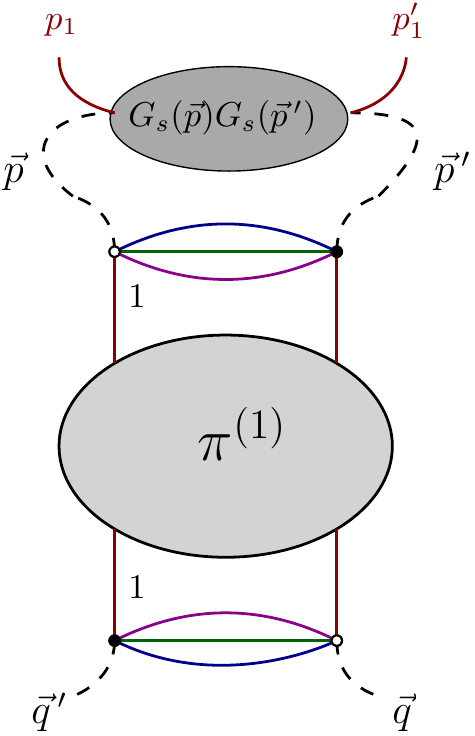}  }}+\vcenter{\hbox{\includegraphics[scale=0.5]{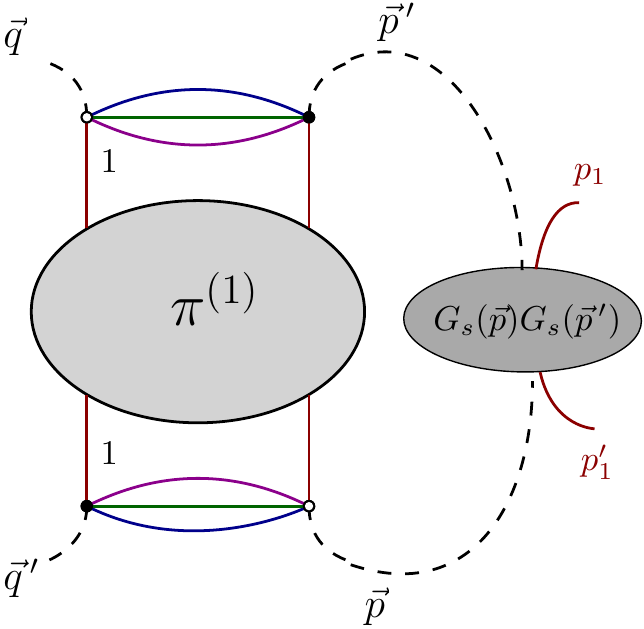}}} \,\,+ \vcenter{\hbox{\includegraphics[scale=0.5]{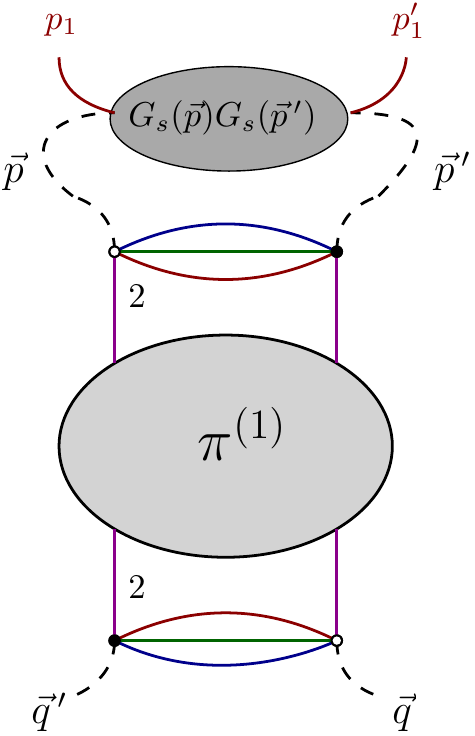}  }}+\cdots\right\}\,.
\end{equation}
In this diagrammatic expression only the first one contribute to the leading order. Moreover, it is easy to see that all other contributions lost an internal face, and then do not contribute at leading order. For instance, this is the case of the last term on the right hand side, in  which we lost an internal faces of color 2. Therefore, taking the limit $p_1\to p_1^\prime$ (we assume that $G_s\in \mathcal{F}(\mathbb{Z})$), and setting $\vec{q}=\vec{q}\,^\prime=\vec{0}$, we find:
\bea
\sum_{\vec{p}_\bot} \frac{1}{2}\Gamma^{(4),\,1}_{s,\vec{0},\vec{0},\vec{0},\vec{0}}\bigg(Z_{-\infty}+\frac{\partial r_{s}(\vec{p}_{\bot})}{\partial p_1^2}\bigg)[G_s(\vec{p}_\bot)]^2=-\frac{\partial}{\partial p_1^2} \left(\Gamma_s^{(2)}(\vec{p}_{\bot})-Z_{-\infty}\vec{p}\,^2\right)\bigg\vert_{\vec{p}=0}\,.\label{ward2}
\eea
This maybe summarize into the following corollary  in terms of $\pi^{(i)}$:
\begin{corollary}\textbf{Zero momenta First WT-identity}\label{FirstWI}
In the symmetric phase, the zero-momenta $4$-point fonction satisfies:
\begin{equation}
 \pi^{(1)}_{00}Z_{-\infty}\mathcal{L}_s=-\frac{\partial}{\partial p_1^2} \left(\Gamma_s^{(2)}(\vec{p}_{\bot})-Z_{-\infty}\vec{p}\,^2\right)\bigg\vert_{\vec{p}=0}\,,
\end{equation}
where we defined the loop $\mathcal{L}_s$ as:
\begin{equation}
\mathcal{L}_s:=\sum_{\vec{p}_\bot}\bigg(1+\frac{\partial \tilde{r}_{s}(\vec{p}_{\bot})}{\partial p_1^2}\bigg)[G_s(\vec{p}_\bot)]^2\,,\qquad r_s=:Z_{-\infty}\tilde{r}_s\,.
\end{equation}
\end{corollary}
\noindent

\section{Melonic structure equations and their consequences}\label{section5}

The recursive definition of the melonic diagrams  which is given in the section \ref{section2}, imply that all correlations functions maybe expressed in terms of the $2$-points function. The explicit form of the correlations functions are obtained in a systematic way, and well established the \textit{melonic structure equations} that we will investigate in this section, as well as their  have  consequences on the renormalization group flow equations. Moreover, we will see that  the structure equations satisfies WT-identities identically. Then, we do not have any additional information coming from Ward identities. The  flow equations improved with structure equations provides effective actions which, satisfy the WI-identities along the flow. The structure equations that we will discuss in the context of the non-perturbative renormalization group have been discussed in recent works \cite{BenGeloun:2011xu}-\cite{Samary:2014oya}.

\subsection{Melonic structure equations and improved $\phi^4$ truncation}

\noindent
The first structure equation concern the self energy (or 1PI $2$-point functions). It takes the form of  the  \textit{closed equation} for self energy \footnote{The rank of the tensors is fixed to $5$, and we denote it by $d$ to clarify the proof(s).}, and state that:
\begin{proposition}\label{lemmatwo}
In the melonic sector, the self energy $\Sigma_s(\vec{p}\,)$ is given by the \textit{closed equation} which takes into account the  effective coupling $\lambda(s)$:
\begin{equation}\label{eq2points}
-\Sigma_s(\vec{p}\,)=2\lambda_rZ_\lambda\sum_{\vec{q}} \left(\sum_{i=1}^d \delta_{p_iq_i}\right)G_s(\vec{q}\,)\,.
\end{equation}
\end{proposition}
\textit{Proof :} Because of the recursive definition of melons, we expect the following structure:
\begin{equation}
-\Sigma_s=\sum_i\vcenter{\hbox{\includegraphics[scale=0.7]{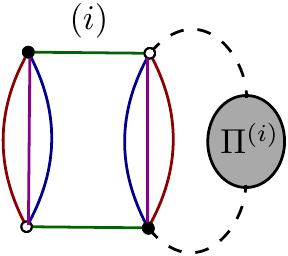} }}=:\sum_i\sigma_s^{(i)}\,.
\end{equation}
This equations may be obtained as follows: In the melonic sector, the expansion of $\Sigma_s$ as a sum of 1PI-Feynman graphs which only involves melonic diagrams. Because of proposition \ref{cormelons}, the external lines are necessarily hooked to the same vertex, and only the configuration of equation \eqref{eq2points} is melonic \cite{Samary:2014tja}. The same argument holds for all diagrams, and the function $\Pi^{(i)}$ is the sum of all heart melonic diagrams in the Feynman expansion. It is not a 1PI function (even if we cut a bridge, the graph remains connected), and with a moment of reflection, one convince oneself that  $\Pi^{(i)}$ is equivalent to the effective propagator $G_s$. 
\begin{flushright}
$\square$
\end{flushright}
 In the same way, for the $4$-points function, we have:
\begin{proposition}\label{prop1} \textbf{ $\phi^4$-structure equation.}
In the melonic sector, the perturbative zero-momenta 1PI four-point contribution $\Gamma^{(4),i}_{s,\vec 0\vec 0;\vec 0\vec 0}$ is given by:
\begin{equation}
\Gamma^{(4),i}_{s,\vec 0\vec 0;\vec 0\vec 0}=2\pi_{00}=\frac{4Z_\lambda\lambda_r}{1+2\lambda_r Z_\lambda \mathcal A_s}\,,
\end{equation}
where $\mathcal A_s$ is defined as:
\begin{equation}
\mathcal A_s=\sum_{\vec p_{\bot}}[G_s(\vec p_{\bot})]^2\,,\,\vec{p}_{\bot} := (0,p_1,\cdots,p_d)\,,
\end{equation}
$G_s(\vec{p})$ being the effective propagator :  $G_s^{-1}(\vec p\,)=Z_{-\infty}\vec p\,^2+m^2+r_s(\vec p\,)-\Sigma_s(\vec p\,)\,.$ Let us recall that $Z_{-\infty}$ and $m_0$ are the counter-terms discarding the UV divergences of the original partition function, the initial conditions in the UV are given such that the classical action contain only renormalizable interactions. \\
\end{proposition}

\noindent
\textit{Proof:} Let us define $4Z_\lambda\lambda_r\Pi$ as the zero momenta melonic $4$-points functions made into  the graphs for which two vertices maybe singularized (i.e. by graphs which are at least of order $2$ in the perturbative expansion). We have\footnote{The notations are similar to the ones used for the previous proof. The context however allows to exclude any confusion.}:
\begin{equation}
2\pi_{00}=:4Z_\lambda\lambda_r(1+\Pi)\,,
\end{equation}
Because of the face connectivity of the melonic diagrams and proposition \eqref{cormelons}, the boundary vertices may be such that the two internal faces of the same color running on the interior of the diagrams building $\Pi$ pass through of them. As a result, we expect the following structure:
\begin{equation}
-4Z_\lambda\lambda_r\Pi=\vcenter{\hbox{\includegraphics[scale=0.5]{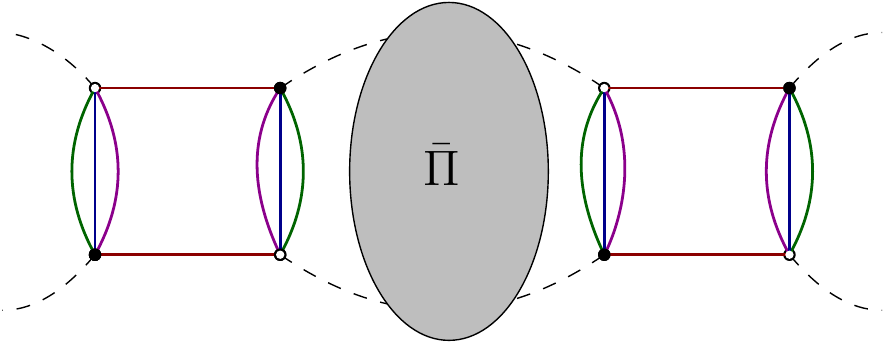} }} 
\end{equation}
where the grey disk is a sum of Feynman graphs. Note that it is the only configuration of the external vertices in agreement with the assumption that $\Pi$ is building with the melonic diagrams. Any other configurations of the external vertices are not melonics. At the lowest order, the grey disk corresponds to propagator lines, 
\begin{equation}
-4Z_\lambda\lambda_r\Pi^{(2)}=8Z_\lambda^2 \lambda_r^2 \mathcal{A}_s\vert_{\lambda_r=0}\equiv\vcenter{\hbox{\includegraphics[scale=0.5]{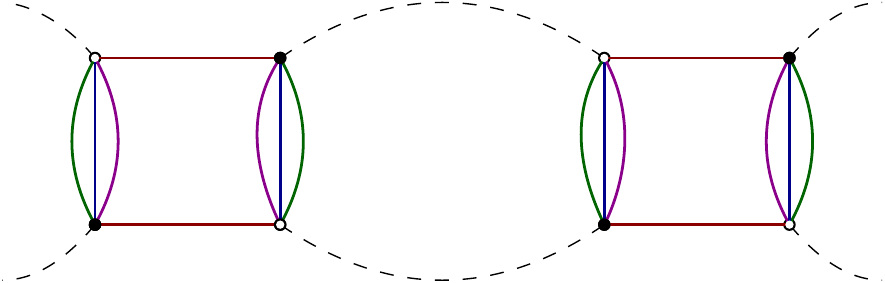} }} \,.
\end{equation}
Note that, as  we explaned  in the section \ref{section2}, the external faces have the same color. Now, we can extract the amputated component of $\bar{\Pi}$, say $\bar{\Pi}^{\prime}$ (which contains at least one vertex, and is irreducible by hypothesis) extracting the effective melonic propagators connected to the  dotted lines linked to $\bar{\Pi}$. We get:
\begin{equation}
-4Z_\lambda\lambda_r\Pi=\vcenter{\hbox{\includegraphics[scale=0.5]{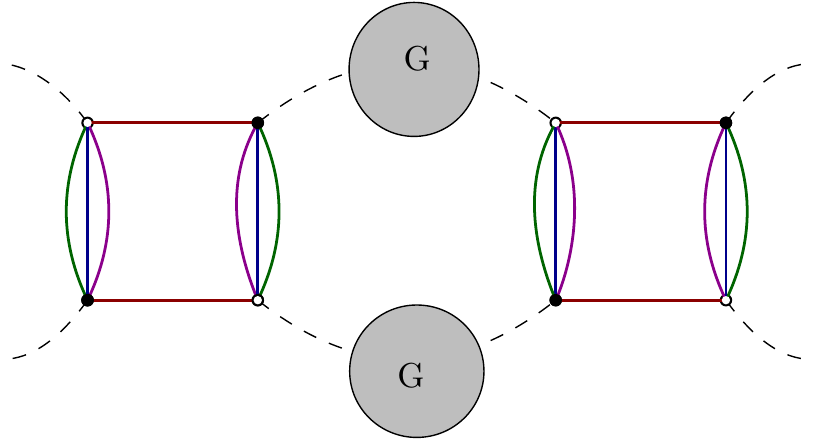} }} +\vcenter{\hbox{\includegraphics[scale=0.5]{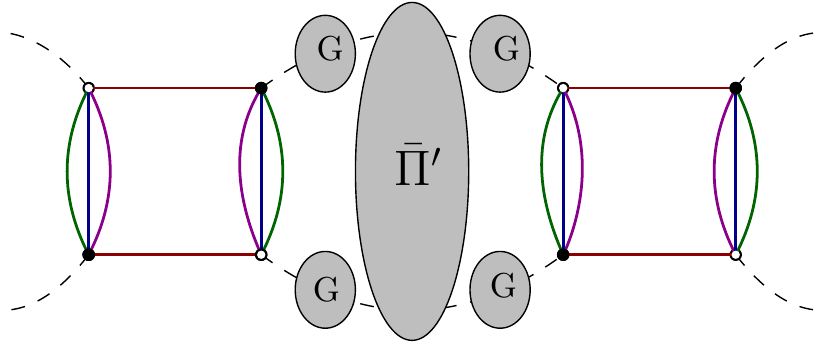} }}\,.
\end{equation}
At first order, $\bar{\Pi}^{\prime}$ is built with a single vertex, and there are only one configuration in agreement with the melonic structure, i.e. maximazing the number of internal faces. The higher order contributions contain at least two vertices, and the argument may be repeated so that the function $\bar{\Pi}^{\prime}$ appears. Finally we deduce the closed relation:
\begin{equation}
\vcenter{\hbox{\includegraphics[scale=0.4]{pirest.pdf} }}=\vcenter{\hbox{\includegraphics[scale=0.4]{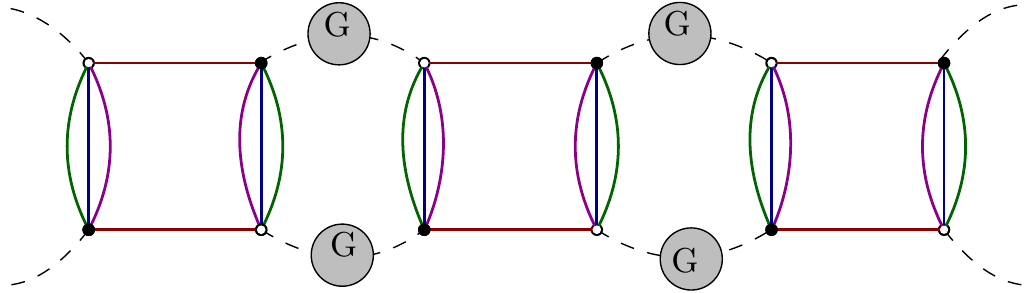} }}+\vcenter{\hbox{\includegraphics[scale=0.45]{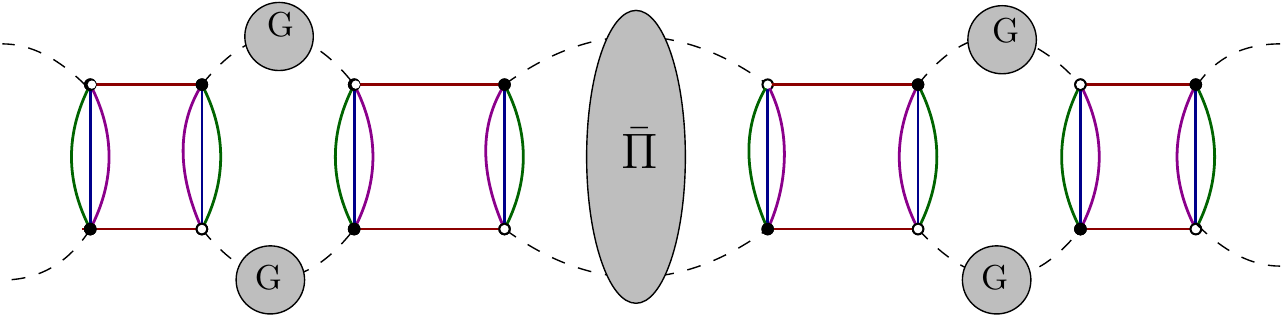} }}\,. \label{recursion1}
\end{equation}
This equation can be solved recursively as an infinite sum
\begin{equation}
-4Z_\lambda\lambda_r\Pi=\vcenter{\hbox{\includegraphics[scale=0.6]{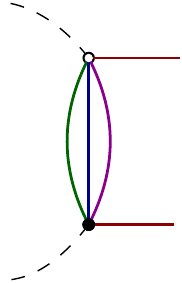} }} \left\{\sum_{n=1}^\infty \left(\vcenter{\hbox{\includegraphics[scale=0.6]{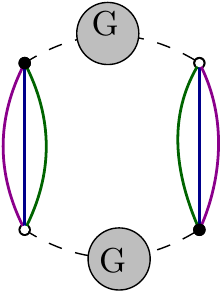} }}\right)^n \right\} \vcenter{\hbox{\includegraphics[scale=0.6]{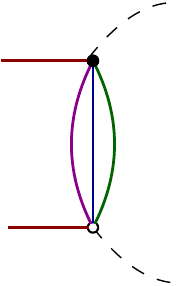} }}\,,
\end{equation}
which can be formally solved as
\begin{equation}
2\pi_{00}= 4Z_\lambda\lambda_r\left(1-\vcenter{\hbox{\includegraphics[scale=0.4]{pimiddle.pdf} }}\right)^{-1}\,.
\end{equation}
The loop diagram $\vcenter{\hbox{\includegraphics[scale=0.4]{pimiddle.pdf} }}$ maybe easily computed recursively from the  definition of melonic diagrams, or directly using Wick theorem  for a one-loop computation with the effective propagator $G$. The result is:
\begin{equation}
\vcenter{\hbox{\includegraphics[scale=0.5]{pimiddle.pdf} }}=-2Z_\lambda \lambda_r \mathcal{A}_s\,,
\end{equation}
and the proposition is proved. \\
\noindent
\begin{flushright}
$\square$
\end{flushright}
 Note this construction can be  also considered in perturbation theory: We can proceed recursively by counting the number of vertices, and show that the power counting remains unchanged at each steps. This is obvious because we have recalled (equation \eqref{melocountinplus}) that divergent degrees is only depended on the number of external lines of the melonic diagrams.\\

These propositions allow to extract the expression of the counter-terms at all orders, and represent an interesting result to show that the wave function renormalization and the $4$-points vertex renormalization are the same. We have:
\begin{proposition}\label{propcounterterms}
Choosing the following renormalization prescription:
\begin{equation}
\Gamma^{(4),1}_{s=-\infty,\vec 0\vec 0;\vec 0\vec 0}=4\lambda_r\,\,;\,\, \Gamma^{(2)}_{s=-\infty}(\vec{p}\,)=m_r^2+\vec{p}\,^2+\mathcal{O}(\vec{p}\,^2)\,,
\end{equation}
where $m_r^2$ and $\lambda_r$ are the renormalized mass and coupling constant; the counter-terms are given by:
\begin{equation}
Z_\lambda=\frac{1}{1-2\lambda_r\mathcal A_{s=-\infty}},\,\,;\,\, Z_{-\infty}=Z_\lambda \,\,;\,\, m^2=m_r^2+\Sigma_{s=-\infty}(\vec{p}=0)\,,
\end{equation}
where $\Sigma_s$ denote the melonic self-energy. 
\end{proposition}
\textit{Proof:} From Proposition \ref{prop1}, we have that:
\begin{equation}
\Gamma^{(4),i}_{s,\vec 0\vec 0;\vec 0\vec 0}=\frac{4Z_\lambda\lambda_r}{1+2\lambda_r Z_\lambda \mathcal A_s}=\frac{4\lambda_r}{Z_\lambda^{-1}+2\lambda_r \mathcal A_s}\,.
\end{equation}
Then, setting $s=-\infty$, we deduce that
\begin{equation}
Z_\lambda^{-1}+2\lambda_r \mathcal A_{-\infty}=1\to Z_\lambda=\frac{1}{1-2\lambda_r\mathcal A_{-\infty}}\,.\label{explicitZlambda}
\end{equation}
We now concentrated our self  on to $Z_{-\infty}$ and $m^2$. Without lost of generality, the inverse of the effective propagator $\Gamma^{(2)}_s$ has the following structure:
\begin{align}
\Gamma_{s=-\infty}^{(2)}(\vec p\,)&=Z_{-\infty}\vec p\,^2 +m^2-\Sigma_{s=-\infty}(\vec p)\\
&=Z_{-\infty}\vec p\,^2 +m^2-\Sigma_{s=-\infty}(\vec 0) -\vec p\,^2\Sigma^\prime_{{s=-\infty}}(\vec 0)+\mathcal{O}(\vec{p}\,^2)\\
&=(Z_{-\infty}-\Sigma_{s=-\infty}^\prime(0))\vec p\,^2+m^2-\Sigma_{s=-\infty}(\vec 0)+\mathcal{O}(\vec{p}\,^2)
\end{align}
with the notation: $\Sigma^\prime(\vec{0}):= \partial \Sigma/\partial p_1^2(\vec{p}=\vec{0}\,)$. Then  from the renormalization conditions, we have :
\begin{equation}
Z_{-\infty}-\Sigma_{s=-\infty}^\prime(0)=1\,\,,\,\, m^2-\Sigma_{s=-\infty}(\vec 0)=m_r^2\,.
\end{equation}
Setting $s=-\infty$ in the closed equation (proposition \ref{lemmatwo}), and by derivating  with respect to $p_1$ for $\vec{p}=\vec{0}$, we get:
\begin{equation}
1-Z_{-\infty}=-2\lambda_r Z_\lambda  \mathcal{A}_{s=-\infty}\,.
\end{equation}
Using the explicit expression for $Z_\lambda$ in  \eqref{explicitZlambda}, we get finally:
\begin{equation}
(1-Z_{-\infty})(1-2\lambda_r  \mathcal{A}_{s=-\infty} )=-2\lambda_r   \mathcal{A}_{s=-\infty} \,\,\to\,\, Z_{-\infty}=Z_\lambda\,.
\end{equation}
\begin{flushright}
$\square$
\end{flushright}
Now, consider the monocolor $4$-points function $\Gamma^{(4),i}_{s,\vec 0\vec 0;\vec 0\vec 0}$. If we replace $Z_\lambda$ by its expression from Proposition \ref{propcounterterms}, we deduce that
\begin{equation}
\Gamma^{(4),i}_{s,\vec 0\vec 0;\vec 0\vec 0}=\frac{4\lambda_r}{1+2\lambda_r  \bar{\mathcal A}_s}\,,
\end{equation}
with the definition: $\bar{\mathcal A}_s:= \mathcal A_s-\mathcal A_{s=-\infty}$. In other words, we have an explicit expression for the effective coupling $\lambda(s):=\frac{1}{4} \Gamma^{(4),i}_{s,\vec 0\vec 0;\vec 0\vec 0}$,
\begin{equation}
\lambda(s)=\frac{\lambda_r}{1+2\lambda_r  \bar{\mathcal A}_s}\,,\label{effectivecoupling}
\end{equation}
from which we get
\begin{equation}
\partial_s\lambda(s)=-\frac{2\lambda_r^2\dot{\mathcal A}_s}{(1+2\lambda_r\Delta \mathcal A_s)^2}=-2\lambda^2(s)\dot{\mathcal A}_s\,.
\end{equation}
In the above relation  we introduce the dot notation $\dot{\mathcal{A}}_s=\partial_s{\mathcal{A}}_s$
\begin{equation}
\mathcal A_s=\sum_{\vec p_{\bot}}\frac{1}{[\Gamma_s^{(2)}(\vec p_{\bot})+r_s(\vec p_{\bot})]^2},\quad 
\dot{\mathcal A}_s=-2\sum_{\vec p_{\bot}}\frac{\dot{\Gamma}_s^{(2)}(\vec p_{\bot})+\dot{r}_s(\vec p_{\bot})}{[\Gamma_s^{(2)}(\vec p_{\bot})+r_s(\vec p_{\bot})]^3}.
\end{equation}
In proposition \ref{propcounterterms} we have investigated the relations between couter-terms i.e. we have considered the melonic equations as Ward identities for $s=-\infty$. Far from the initial conditions, the Taylor expansion of the $2$-point function $\Gamma_s^{(2)}(\vec{p}\,)$ is written  as:
\begin{equation}
\Gamma_s^{(2)}(\vec p\,)=m_r^2+(\Sigma_s(\vec 0\,)-\Sigma_0(\vec 0\,))+(Z_{-\infty}-\Sigma_s'(\vec 0))\vec p\,^2+\mathcal{O}(\vec{p}\,^2)\,.
\end{equation}
We call the "physical" or \textit{effective} mass parameter $m^2(s)$ the first term in the above relation:
\begin{equation}
m^2(s):=m_r^2+(\Sigma_s(\vec 0\,)-\Sigma_0(\vec 0\,)),\,
\end{equation}
while the coefficient $Z_{-\infty}-\Sigma_s'(\vec 0)$ is the effective wave function renormalization and is denoted by $Z(s)$ i.e.
\begin{equation}
Z(s):=Z_{-\infty}-\Sigma_s'(\vec 0)\,.
\end{equation}
Now let us consider the closed equation given in proposition \ref{lemmatwo}. By  derivating  with respect to $p_1$ and by taking $\vec{p}=\vec{0}$, we get:
\begin{equation}
Z-Z_{-\infty}=-2\lambda_r Z_\lambda \sum_{\vec{p}_\bot} G^2_s(\vec{p}_\bot)(Z+r'_s(\vec{p}_\bot))\,.
\end{equation}
Using equation \eqref{effectivecoupling}, we can express $\lambda_r Z_\lambda$ in terms of the effective coupling $\lambda(s)$, and we get:
\begin{equation}
(Z-Z_{-\infty})(1-2\lambda(s)\mathcal{A}_s)=-2\lambda(s)\left(Z\mathcal{A}_s+\sum_{\vec{p}_\bot} G^2_s(\vec{p}_\bot)r'_s(\vec{p}_\bot)\right)\,,
\end{equation}
Then we come to the following relation
\begin{equation}
Z=Z_{-\infty}\left(1-2\lambda(s)\mathcal{L}_s\right)\,.\label{eqZ}
\end{equation}
At this stage, without all confusion let us clarify that:       $Z_{-\infty}$ is the wave function counter-term i.e, whose divergent parts cancels the loop divergences, and whose finite part depend on the renormalization prescription. $Z(s)$ however is  fixing to be $1$ for $s=-\infty$ from our renormalization conditions. \\

\noindent
Our goal for this section require the structure equation for melonic $6$-points functions given in the following proposition:
\begin{proposition} \textbf{$\phi^6$-structure equation. }\label{lemmasix}
In the melonic sector, the $6$-point functions $\Gamma_{s,\cdots}^{(3,3)}$ expanded in the symmetric phase have the following structure:
\begin{equation}\label{lefthand}
\Gamma_{s,\vec{p}_2\vec{p}_4\vec{p}_6;\vec{p}_1\vec{p}_3\vec{p}_5}^{(3,3)}=\sum_{i=1}^d \Gamma_{s,\vec{p}_2\vec{p}_4\vec{p}_6;\vec{p}_1\vec{p}_3\vec{p}_5}^{(3,3),\,i}\,,
\end{equation}
where :
\begin{equation}
\Gamma_{s,\vec{p}_2\vec{p}_4\vec{p}_6;\vec{p}_1\vec{p}_3\vec{p}_5}^{(3,3),\,i}=:\pi^{(i)}_{3,p_{1i}p_{3i}p_{5i}}\delta_{p_{1i}p_{6i}}\delta_{p_{5i}p_{4i}}\delta_{p_{3i}p_{2i}}\delta_{\vec{p}_{\bot_i1}\vec{p}_{\bot_i{2}}}\delta_{\vec{p}_{\bot_i3}\vec{p}_{\bot_i{4}}}\delta_{\vec{p}_{\bot_5}\vec{p}_{\bot_i{6}}}\,+\,\perm(\vec{p}_1,\vec{p}_3,\vec{p}_5)\,.
\end{equation}
\end{proposition}
\noindent
\textit{Proof } Consider the melonic contribution to $ \Gamma_{s,\vec{p}_2\vec{p}_4\vec{p}_6;\vec{p}_1\vec{p}_3\vec{p}_5}^{(3,3)}$. From proposition \ref{cormelons}, it follows that external lines are hooked to three vertices of the same type. Moreover, they share the three external faces of the same color running in the heart graph. Because these heart external faces have the same color, they defined $d$ components, that we call $ \Gamma_{s,\vec{p}_2\vec{p}_4\vec{p}_6;\vec{p}_1\vec{p}_3\vec{p}_5}^{(3,3),\,i}$, which have the following structure:
\begin{equation}
 \Gamma_{s,\vec{p}_2\vec{p}_4\vec{p}_6;\vec{p}_1\vec{p}_3\vec{p}_5}^{(3,3),\,i}=\vcenter{\hbox{\includegraphics[scale=0.5]{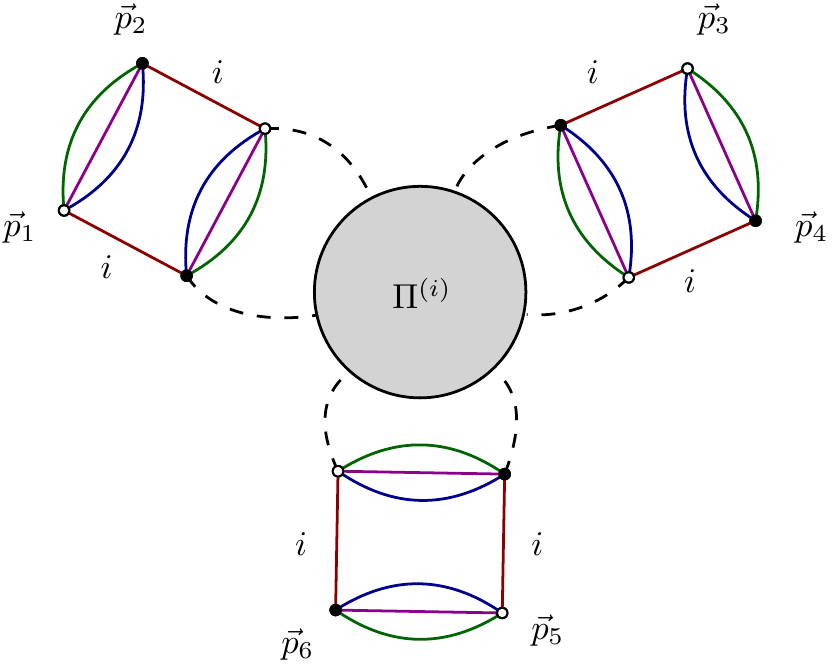} }}+\perm(\vec{p}_1,\vec{p}_3,\vec{p}_5)\,,
\end{equation}
where $\Pi^{(i)}$ is of order zero (i.e. its proper Feynman expansion start without vertices). As for the $4$-points function we can discard the external propagators, and we get 
\begin{equation}\label{satttt}
\vcenter{\hbox{\includegraphics[scale=0.5]{structuresixpoints.pdf} }}=\vcenter{\hbox{\includegraphics[scale=0.5]{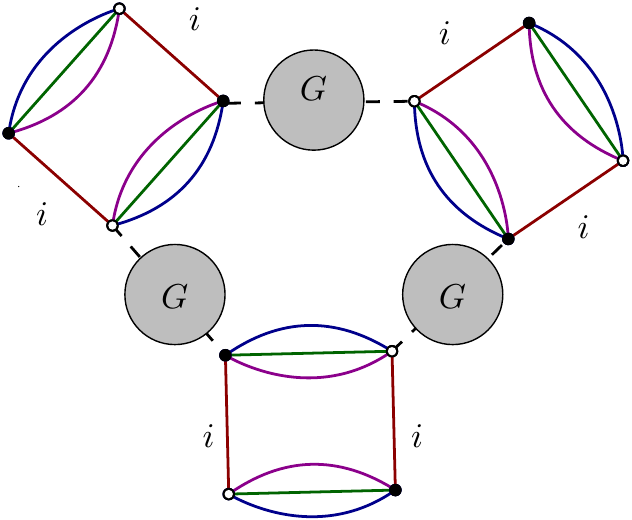} }}\,+\,\vcenter{\hbox{\includegraphics[scale=0.5]{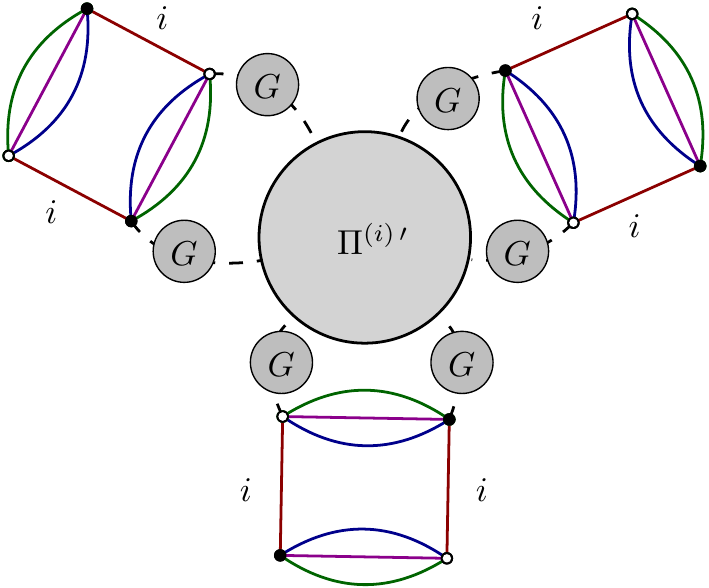} }}\,.
\end{equation}
Without all confusion we use the notation $\Pi^{(i)}$ in the expression \eqref{satttt}, which is  already used for the self energy. It is clear that it is not the same function.
The recursion relation is equivalent to the equation  \ref{recursion1}  for the $4$-point function.    Indeed, $\Pi^{(i)\,\prime}$ is equivalent to the amputated of the function $\Pi^{(i)}$, and the resulting equation maybe solved following from the same recursive method like the result which leads to the $4$-point function. We get ($\perm(\vec{p}_1,\vec{p}_3,\vec{p}_5)$ is omitted):
\begin{equation}
 \Gamma_{s,\vec{p}_2\vec{p}_4\vec{p}_6;\vec{p}_1\vec{p}_3\vec{p}_5}^{(3,3),\,i}=\vcenter{\hbox{\includegraphics[scale=0.5]{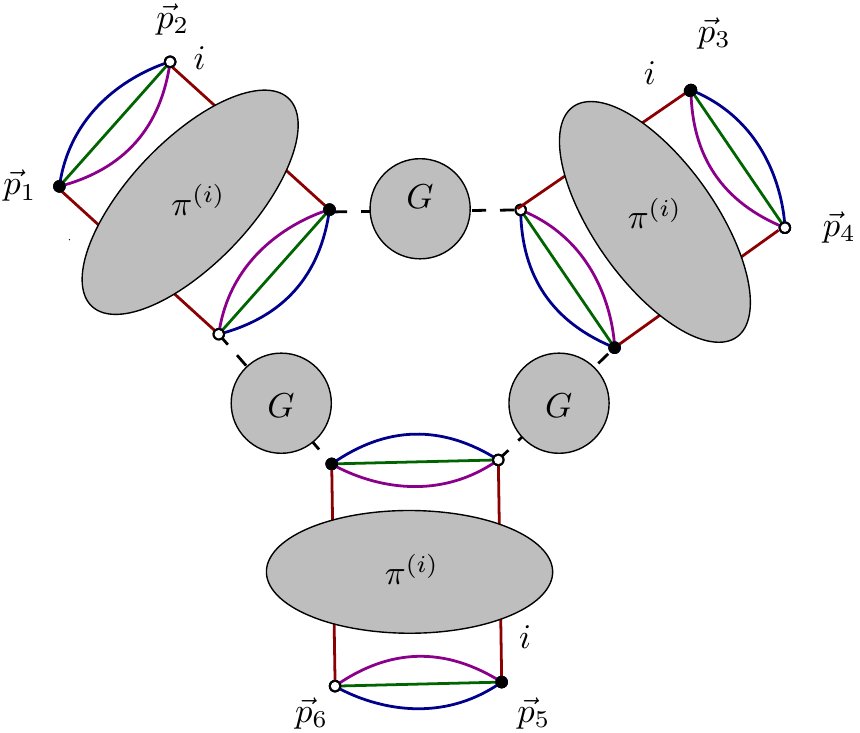} }}\,\to \, \pi^{(i)}_{3,p_{1i}p_{3i}p_{5i}}=\vcenter{\hbox{\includegraphics[scale=0.5]{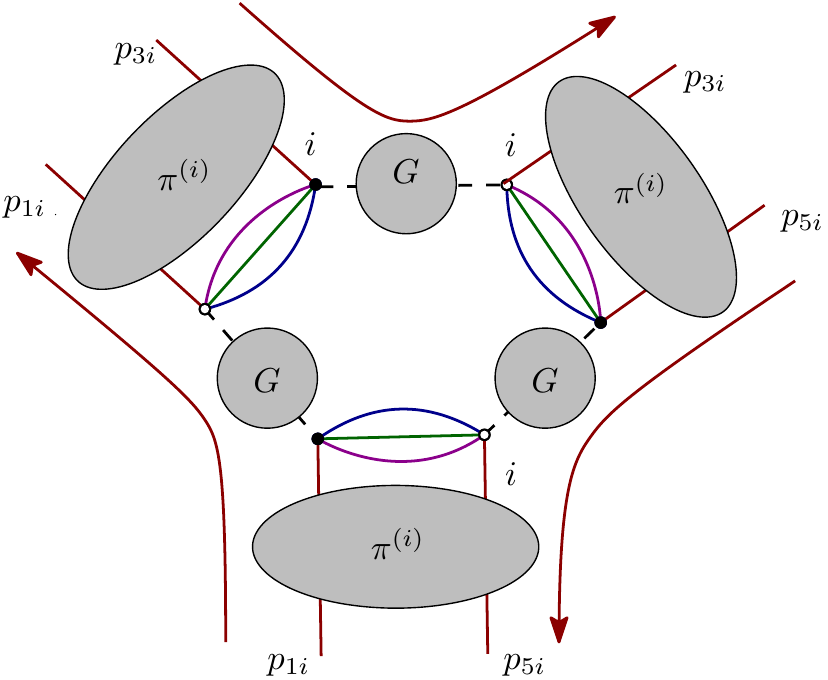} }}\,.\label{sixpointstructure}
\end{equation}
The function $\pi_3^{(i)}$ may be directly identified in the left hand side of the expression \eqref{lefthand}. The proposition in then proved!
\begin{flushright}
$\square$
\end{flushright}
Ler us remark that we only need the zero momentum value $\pi^{(i)}_{3,000}=:\pi_3$. Its explicit expression can be easily deduced from the structure of the graph on \eqref{sixpointstructure}, up to a purely numerical number which can be straightforwardly computed from Wick theorem. We get:
\begin{equation}
\pi_3(s)=16\lambda^3(s) \sum_{\vec{p}_\bot} \left[G_s(\vec{p}_\bot)\right]^3=16Z^3(s)\bar{\lambda}^3(s)e^{-2s}\bar{\mathcal{A}}_{3s}\,.\label{exppi3}
\end{equation}

\begin{remark}

(i)  Note that the dimension of $\pi_3(s)$ given in  \eqref{exppi3} is  $[\pi_3(s)]=-2$. Our final goal is to use the structure equations given on the propositions \ref{prop1}, \ref{lemmatwo}, \ref{lemmasix}  to solve the hierarchies equations obtained by expanding the exact flow equation \eqref{wetterich} in its irreducible parts. More precisely, the structure equations express all the 1PI functions in term of the quartic melonic coupling and the self energy, which introduce a natural cutoff in the hierarchies equations around the marginal coupling. The derivation of the corresponding equations is given in the proposition \ref{flowstructure}.

(ii)  We have to make the following important remarks. As announced at the beginning of section \ref{section3}, there are  equivalence between WT-Identities and melonic structure equations. Indeed, we do not introduce any  additional constraint on the flow equation which is  obtained just from the structure equations. However, at least in the melonic sector, the Ward identities are completely redundant  with respected to the structure equations. To be more precise, let us consider the WI given from Proposition \ref{FirstWI}, 
\begin{equation}
 \pi^{(1)}_{00}Z_{-\infty}\mathcal{L}_s=-\frac{\partial}{\partial p_1^2} \left(\Gamma_s^{(2)}(\vec{p}_{\bot})-Z_{-\infty}\vec{p}\,^2\right)\bigg\vert_{\vec{p}=0}\,.
\end{equation}
Now, $\pi^{(1)}_{00}=2\lambda(s)$ and the last term is  $Z(s)-Z_{-\infty}$.  We can  show that this equation is  reduced to \eqref{eqZ}. In appendix \ref{AppA} we show the same equivalence for $6$-point structure equations, and state that it is true for higher irreducible functions. 

(iii) The second remark concern the notion of \textit{local interactions}. This notion have been briefly discussed in the introduction. Obviously our interactions are not local on the  $U(1)^5$ group. However, the locality  for tensorial interactions and \textit{traciality} is given through a list of references in the litterature. See   \cite{Carrozza:2013mna}-\cite{Geloun:2013saa} and references therein. For our purpose, a local interaction is a sum of connected melonic tensorial invariants as they have been defined in section \eqref{section2}.
  For our purpose, and in accordance with standard conventions, we call \textit{local potential approximation} any effective average action expanded as a sum of connected tensorial interaction.

(iv) Finally, before give  the key proposition of this section, we  introduce the notion of \textit{ renormalized sums}. All the equations that we will consider are of the form:
\begin{equation}
\mathcal{S}_{l,n}(s,p)=\sum_{\vec{q}_\bot} G\,^n_s(\vec{q}\,) \dot{r}_s^{l}(\vec{q}\,)\bigg\vert_{q_1=p}\,,
\end{equation}
where  $l=\{0,1\}$. To built the renormalized sum, we extract the scaling in $s$ and the global factor $Z(s)$ coming from the renormalization of $\Gamma^{(2)}_s$ : $\bar{\Gamma}^{(2)}_s=Z^{-1}{\Gamma^{(2)}}_s$. To this end, we assume that $r_s$ share a factor $Z$, and has canonical dimension $2$. Then, the renormalized sum ${\mathcal{S}}_{\{i_l\},\{j_l\},k,n}(s)$ corresponds to:
\begin{equation}
\mathcal{S}_{l,n}(s,p)=Z^{l-n}(s)e^{2(l-n+2)s}\bar{\mathcal{S}}_{l,n}(s,p).\,
\end{equation}
Note that ${\mathcal{S}}_{0,n}\equiv \mathcal{A}_{ns}$   and  as  $\mathcal{A}_{ns}$ we simply denote by $\mathcal{S}_{l,n}$ the sum for $p=0$ in the rest of the paper. In the same way, $\mathcal{S}_{l,n}^\prime$ will be $\partial \mathcal{S}_{l,n}(p=0)/\partial p^2$. \\
\end{remark}
\noindent
We are now in position to enunciate the key proposition of this section:
\begin{proposition}\label{flowstructure}
\textbf{(Improved $\phi^4$ truncation)} In the UV sector $ 1\ll e^s=k\ll \Lambda$, the exact flow equations for the dimensionless-renormalized essential and marginal couplings, i.e. for mass, wave function and quartic melonic interaction are given by:
\begin{equation}
\left\{
    \begin{array}{ll}
       \beta_m &= -(2+\eta_s)\bar{m}^2-2d\bar{\lambda} \bar{\mathcal S}_{1,2} \\
       \eta_s&=\,\,\,\,4\bar{\lambda}^2 \bar{\mathcal{S}}_{0,2}^\prime  \bar{\mathcal{S}}_{1,2}-2\bar{\lambda} \bar{\mathcal{S}}_{1,2}^{\prime}\\
        \beta_\lambda &=-2\eta_s\bar{\lambda}+4\bar{\lambda}^2 \bar{\mathcal S}_{1,3}-16\bar{\lambda}^3 \bar{\mathcal S}_{1,2}\bar{\mathcal{A}}_{3s}
    \end{array}
\right. 
\end{equation}
\end{proposition}
\noindent
\textit{Proof:}  By taking the partial derivative of  the flow equation \eqref{wetterich} with respect to $\bar{M}_{\vec{p}}$ and to $M_{\vec{p}}$, we get:
\begin{equation}
\partial_s\Gamma_s^{(2)}(\vec p)=-\sum_{\vec q}\Gamma^{(4)}_{s,\vec p\vec q;\vec p,\vec q}\frac{\dot r_s(\vec q\,)}{[\Gamma_s^{(2)}(\vec q\,)+r_s(\vec q\,)]^2}=-\sum_{i}\sum_{\vec q}\Gamma^{(4),i}_{s,\vec p\vec q;\vec p,\vec q}\frac{\dot r_s(\vec q\,)}{[\Gamma_s^{(2)}(\vec q\,)+r_s(\vec q\,)]^2}\,.
\end{equation}
The leading order contractions with the effective propagator $\frac{\dot r_s(\vec q)}{[\Gamma_s^{(2)}(\vec q)+r_s(\vec q)]^2}$ only concern \textit{tadpole contractions} over the same external vertices of $\Gamma^{(4),i}_{s,\vec p\vec q;\vec p,\vec q}$ (see Figure \ref{contractionsmelo}).

\begin{center}
\includegraphics[scale=0.7]{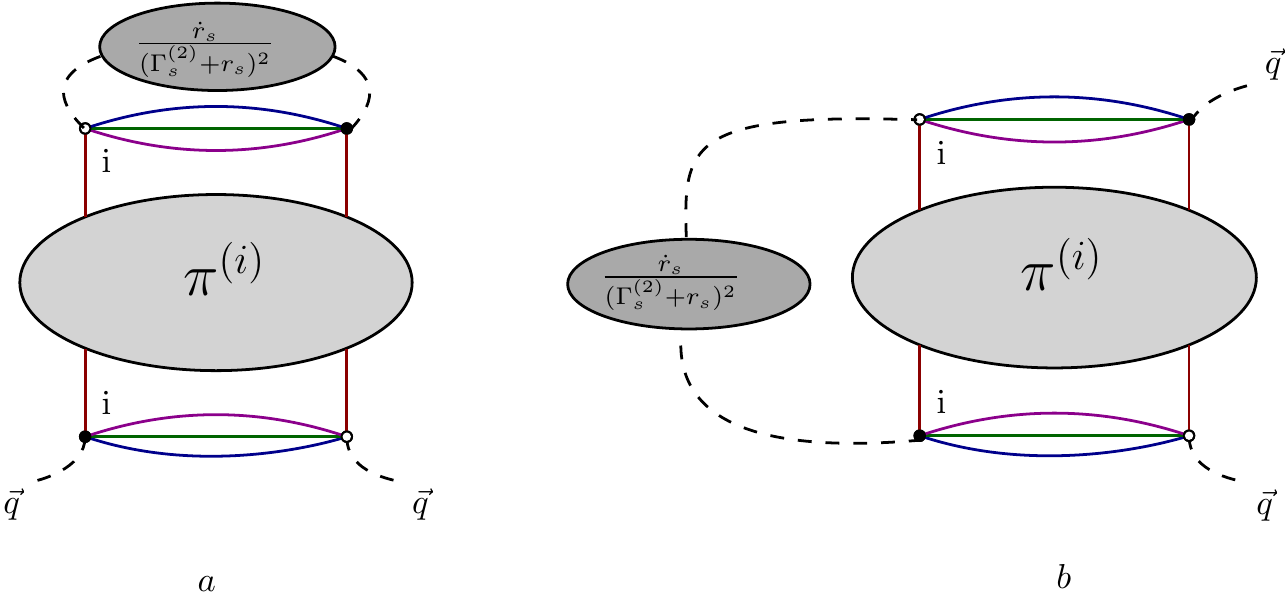} 
\captionof{figure}{Contractions of the effective propagator $\dot{r}G^2_s$ with the effective $4$-point function. On the left (a) the contraction is melonic and create $d-1$ internal faces, whereas on the right (b) it create only a single red internal face. Then, such a contribution maybe discarded in the UV regime.} \label{contractionsmelo}
\end{center}

\noindent
Then, keeping only the melonic contributions, we find   (using the notations of section \ref{section3}) that:
\begin{equation}
\partial_s\Gamma_s^{(2)}(\vec p )=-\sum_{i}\sum_{\vec q}\gamma^{(4),i}_{s,\vec p\vec q;\vec p,\vec q}\frac{\dot r_s(\vec q \,)}{[\Gamma_s^{(2)}(\vec q \,)+r_s(\vec q\,)]^2},\quad \gamma^{(4),i}_{s,\vec p\vec q;\vec p,\vec q}=\pi^{(i)}_{s,p_ip_i}\mathcal{W}^{(i)}_{\vec p\vec p\vec q\vec q}\,.\label{floweq2p}
\end{equation}
The $4$-point function $\pi^{(i)}_{s,p_ip_i}$ may be computed following exactly the same strategy as for proposition \ref{prop1}, then:
\begin{equation}
\pi^{(i)}_{s,p_ip_i}=\frac{2\lambda_r}{1+2\lambda_r(\mathcal{S}_{0,2}(p_i)-\mathcal{A}_{-\infty})}\,.
\end{equation}
Setting $\vec{p}=\vec{0}$ in the flow equation \ref{floweq2p} and because $\Gamma_s^{(2)}(\vec{0})$ is identify to  $m^2(s)$, we deduce that
\begin{equation}
\partial_s m^2(s)=-2d\lambda(s)\bar{\mathcal S}_{1,2}\,,\label{massflow}
\end{equation}
and the equation for $\beta_m:=\dot{\bar{m}}^2$ follows straightforwardly from:
\begin{equation}
\partial_s\bar{m}^2+(2+\eta_s)\bar{m}^2=:\beta_m+(2+\eta_s)\bar{m}^2=e^{-2s}Z(s)^{-1}\partial_sm^2\,. 
\end{equation}
Now, consider the first derivative of the flow equation \ref{floweq2p} with respect to $p_1^2$ evaluated at $\vec{p}=0$. In the symmetric phase, the derivative of the $2$-point function evaluated at zero momenta is equal to the effective wave function renormalization:
\begin{equation}
\frac{d\Gamma_s^{(2)}}{dq_1^2}(\vec{q}=\vec{0})=Z(s)\,,\label{defZ}
\end{equation}
and we get:
\begin{align}
\partial_s\frac{d\Gamma_s^{(2)}(\vec 0)}{dp_1^2}&=-\frac{d\pi_{s,00}^{(1)}}{dp_1^2}\mathcal S_{1,2}-\pi_{s,00}^{(1)}\frac{d}{dp_1^2}\sum_{\vec q_\bot}\frac{\dot r_s(\vec q_\bot)}{[\Gamma_s^{(2)}(\vec q_\bot)+r_s(\vec q_\bot)]^2}\Bigg|_{p_i=q_i=0}\cr
&=4\lambda^2(s) \mathcal{S}_{0,2}^\prime\mathcal S_{1,2}-2\lambda(s) \mathcal{S}_{1,2}^{\prime}\,.
\end{align}
Then, extracting the renormalized part, we can derive the two first equations of the proposition. For the last flow equation, using \ref{floweq2p}  we get
\begin{align}
\nonumber\partial_s\Gamma^{(4)}_{s,\vec 0,\vec 0,\vec 0,\vec 0}=-\sum_{\vec p}\dot r_s(\vec p\,) G^2_s(\vec p\,)&\Big[\Gamma^{(6)}_{s,\vec p,\vec 0,\vec 0,\vec p,\vec 0,\vec 0}-2\sum_{\vec p\,'}\Gamma^{(4)}_{s,\vec p,\vec 0,\vec p\,',\vec 0}G_s(\vec p\,')\Gamma^{(4)}_{s,\vec p\,',\vec 0,\vec p,\vec 0}\Big]\\
&+2\sum_{\vec p}\dot r_s(\vec p\,) G^3_s(\vec p\,)[\Gamma^{(4)}_{s,\vec p,\vec 0,\vec p,\vec 0}]^2\,. \label{flowfour}
\end{align}
\noindent
Setting $\vec{p}=\vec{0}$ in the flow equation \eqref{flowfour} and by using Lemma \ref{lemmasix}, the definition of the effective coupling $\lambda(s)$ (\eqref{effectivecoupling}) and the definition (equation) \eqref{eqfour} of $\pi^{(i)}_{pp^\prime}$, and keeping only the leading order contractions as for the computation of the flow equation for $2$-point observables, we obtain:
\begin{align}
\nonumber 4\partial_s\lambda&=-\sum_{\vec p}\dot r_s(\vec p) G^2_s(\vec p)\Gamma^{(3,3),\,1}_{s,\vec p,\vec 0,\vec 0,\vec p,\vec 0,\vec 0}+4(\pi_{s,00}^{(1)})^2\sum_{\vec p_\bot}\dot r_s(\vec p_\bot) G^3_s(\vec p_\bot)\\
&=-2\mathcal{S}_{1s}\pi^{(1)}_{3,000}+4(\pi_{s,00}^{(1)})^2\mathcal{S}_{3s}\,.\label{flowfour2}
\end{align}
where the first factor $2$ comes from the counting of independent leading order contractions: For a melonic contraction, the end points contracting the $\dot r_s G^2_s$ are necessarily on the same boundary vertex. Then, we have $4$ such diagrams, corresponding to the permutations of the remaining boundary variables. 
\begin{flushright}
$\square$
\end{flushright}

\subsection{Higher structure equations}

The quantity that we called $\pi_3(s)$ is identified to the  effective interaction of valence six. The combinatoric of the effective vertex correspond to the following tensorial interaction:
\begin{equation}
\mathcal{V}^{(3,3)\,,(i)} =\pi_3(s) \,\times\,\vcenter{\hbox{\includegraphics[scale=0.6]{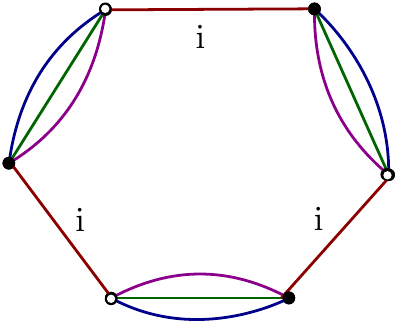} }}\,,
\end{equation}
in virtue of why it legitimate to call $\lambda_6(s)$ the effective coupling for this interaction. Because of the explicit expression \eqref{exppi3}, we deduce the beta function:
\begin{equation}
\beta_{6}=24\beta_\lambda\bar{\lambda}^2(s)\bar{\mathcal{A}}_{3s}+16\bar{\lambda}^3(s)\partial_s\bar{\mathcal{A}}_{3s}\,.
\end{equation}
\noindent
This method maybe extended for effective vertices with arbitrary valence. Note that we do not generate all melonic vertices from our initial conditions for the effective action. As an example, there are no contraction of quartic melons which allows to reproduce the effective vertex:
\begin{equation}
\includegraphics[scale=0.7]{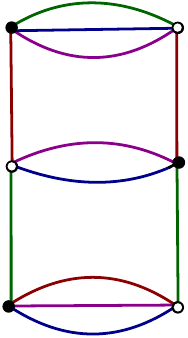}\,.
\end{equation}
The effective coupling  $\pi_N(s)$ for local interaction of valence $2N$ may be computed following the same method as $\pi_3$, and we have the following result:
\begin{proposition}\label{prophigher}
In the melonic (UV) sector, the effective coupling $\pi_N(s)$ satisfy:
\begin{equation}
\pi_N(s)=(2\lambda)^N\sum_{v=1}^{N-2}\sum_{\mathcal{T}_v^{c}}(2\lambda)^{v-1} \prod_{n=1}^v \mathcal{A}_{c_n}\prod_{k=1}^v(n_k-1)!\,,
\end{equation}
where $n_k$ is the number of cordinations of degre $k$, and $\prod_{k=1}^v(n_k-1)!$ corresponds to the number of independent Wick contraction ensuring the connectivity of the graph.
\begin{equation}
\mathcal{A}_{n}:=\sum_{\vec{p}_\bot} \left[G_s(\vec{p}_\bot)\right]^n\,,\qquad \mathcal{A}_{2}\equiv\mathcal{A}\,,
\end{equation}
and $\mathcal{T}_v^{c}$ are trees with mono-colored edges of color $c$, with $N$ $c$-colored external edges and boundary nodes with at least two external edges. 
\end{proposition}
\noindent
Note that this expression involve the canonical dimension of $\pi_N$. Indeed, $\lambda$ is dimensionless,  we have  to consider only the first term of the expansion: $[\pi_N]=[\mathcal{A}_{Ns}]=2(2-N)$.\\

\noindent
\textit{Proof:} For convenience, the proof of this proposition require an appropriate notation for graphs. From  now, we denote by \includegraphics[scale=0.7]{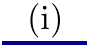}  the elementary melon with index of  color $i$, that is:
\begin{equation}
\vcenter{\hbox{\includegraphics[scale=1]{melonbubble2.pdf} }}=\vcenter{\hbox{\includegraphics[scale=1]{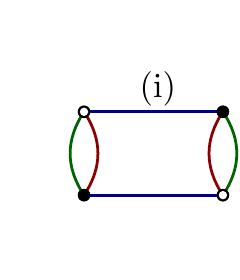} }}
\end{equation}

\noindent
Moreover, to each loops made with a chain of $N$ effective propagator, we have considered only the structure equations and we draw a point with $N$ colored external lines. For instance:
\begin{equation}\label{sam222}
\vcenter{\hbox{\includegraphics[scale=1]{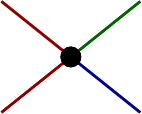} }}\,,
\end{equation}
the graph given in figure  \eqref{sam222} corresponds to  a $8$-point function with two external vertices of color red, one of color blue and one of color green, each ‘‘arc" or \textit{corner} being one of the $N$ effective propagator in the loop. For the rest, we denote by \textit{edges} and \text{nodes} the lines and vertices of this representation, keeping the terminology vertex and line for the standard diagrams. \\
\noindent
Using the above convention, we can establish the proof of our proposition. Remark that a general melonic effective vertex receive many contributions. To proceed step by step, we  are setting $N=4$, which is the first non-trivial case. Obviously, from the same argument used for building the $N=3$ functions, we have the contributions 
\\
\begin{equation}
\vcenter{\hbox{\includegraphics[scale=1]{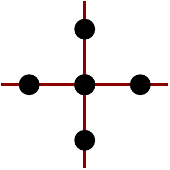} }}\,.
\end{equation}
 Indeed, diagrams contributing to this function have power counting $\omega=4-8=-4$, and it is easy to see that the configuration:
\begin{equation}
\vcenter{\hbox{\includegraphics[scale=1]{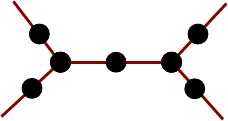} }}\,,\label{sum1}
\end{equation}
Is also a leading order graph. Indeed, when we merge two diagrams $\mathcal{G}_m$ and $\mathcal{G}_n$ to a diagram $\mathcal{G}_N$  following the rule pictured on \eqref{sum1}, the power counting of $\mathcal{G}_N$ is:
\begin{equation}
\omega(\mathcal{G}_N)=\omega(\mathcal{G}_n)+\omega(\mathcal{G}_m)-F_{\mathcal{G}_n\cap\mathcal{G}_m}\,,
\end{equation}
where the last term comes from the fact that the internal faces are counted twice. It is clear that optimal counting  gives $F_{\mathcal{G}_n\cap\mathcal{G}_m}=0$, but it is exactly the case of diagram \eqref{sum1}. Note that we do not create any new common internal face. Then, the power counting of $6$-point melonic graphs being $-2$, the sum match with the power counting for melonic $8$-point graphs. The full $8$-point melonic function\footnote{Strictly speaking, it is the red colored component of the $8$-melonic function.}, that we designate with a white circle, is  weighted sum of these contributions:
\begin{equation}
\vcenter{\hbox{\includegraphics[scale=1]{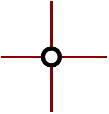} }}=\frac{1}{4!}\left\{K_1\,\vcenter{\hbox{\includegraphics[scale=1]{fourpointnew2.pdf} }}+K_2\,\vcenter{\hbox{\includegraphics[scale=1]{fourpointnew3.pdf} }}\right\}\,,\label{pi4}
\end{equation}
where $K_1$ and $K_2$ are combinatorial coefficients which can be easily computed counting the number of melonic contractions at lower order in perturbation theory. For the first diagram, we have $(4!)^2$ ways to contract external fields (one $4!$ for $T$'s and one for $\bar{T}$'s). Moreover, each vertex has an axial symmetry, which generate a factor $2$ per vertex. Finally, a factor $1/4!$ comes from the exponential expansion, leading to $K_1=4! 2^4$. In the same way, for the second diagram, we have $5$ manners to choose the bridge vertex, $(4!)^2$ coming from the contractions of external fields, and $1/5!$ coming from the exponential. Adding the $2^5$ dues to the axial symmetry, we get $K_2=4! 2^5$. Note that $4!$ is equal to  the cardinality of $\perm(\vec{p}_1,\vec{p}_2,\vec{p}_3,\vec{p}_4)$, and have to be discarded for the definition of the effective vertex $\pi_4$, as for the definition of $\pi_3$ (it corresponds to the factor $1/4!$ in front of \eqref{pi4}).\\

\noindent
The same argument can be easily generalized : Any mono-colored tree build with this rule is melonic. Moreover, it is easy to see that any loop created with colored lines provides a sub-leading diagram : Any created loop increases the number of internal faces by $1$, which do not compensate the number of added internal lines with respect to the tree configuration (at least two). As a result, this contributions are mono-colored trees as:
\begin{equation}
\vcenter{\hbox{\includegraphics[scale=1]{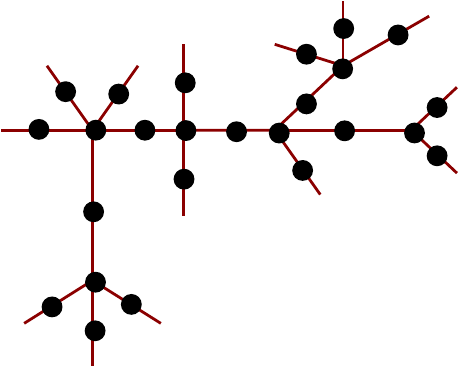} }}\,,
\end{equation}
and the equation for $\pi_N$ may be written as a sum over mono-colored trees:
\begin{equation}
\pi_N=\lambda^N\sum_{v=1}^{N-2}\sum_{\mathcal{T}_v^{c}}K_{\mathcal{T}_v^{c}}\lambda^{v-1} \prod_{n=1}^v \mathcal{A}_{c_n}\prod_{k=1}^v(n_k-1)!\,,
\end{equation}
where the bound number $N-2$ comes from the fact that the maximal length for a tree with $N$ external lines is $N-2$ and the $\mathcal{T}_v^{c}$ are mono-colored trees with $v$ nodes, edges of color $c$ and $N$ opening lines (except the external edges, a tree have $v-1$ edges). Moreover, note that the trees on which we sums have another important property: Their boundary nodes (i.e. nodes hooked to the rest of the tree with a single colored edge) have at last \textit{two external edges}. The $K_{\mathcal{T}_v^{c}}$, indexed by trees are combinatorial factor taking into account symmetries of the trees. The last term is the product over all loops functions $ \mathcal{A}_{c_n}$ for coordination number (number of hooked colored edges) $c_n$ at the node $n$. Note that the $c_n$ satisfy the constraint: $\sum_n c_n=N+v$. The factors $ K_{\mathcal{T}_v^{c}}$ count the number of way providing a given tree. To compute it, we proceed as for the $8$-point function and consider the lower term in the perturbative expansion. Firstly, we have a factor $1/(N+v-1)!$ coming from the exponential (we have $N+v-1$ red vertices). On the other hand, we have $(N+v-1)(N+v-2)\cdots(N+1)$ choices for the bridge vertices, providing a factor $(N+v-1)!/N!$, and the factor $(N!)^2$ coming from the possibles contractions of the remaining vertices with external fields. Finally, the axial symmetry provides a factor $2^{N+v-1}$, and we have to add an additional $1/N!$ factor to remove the cardinality of $\perm(\vec{p}_1,\cdots,\vec{p}_N)$. The resulting factor $K_{\mathcal{T}_v^{c}}$ is then:
\begin{equation}
K_{\mathcal{T}_v^{c}}=\frac{1}{N!}\frac{1}{(N+v-1)!}\frac{(N+v-1)!}{N!}\times (N!)^2 \times 2^{N+v-1}=2^{N+v-1},\,
\end{equation}
and this does not depend on the structure  of the tree, but just  only  depend on their opening lines and the number of nodes. 

\begin{flushright}
$\square$
\end{flushright}

 The notations used in the proof of proposition \eqref{prophigher} involve the set of  diagrams which are re-summed and being the version of well known diagrams in \textit{intermediate field representation} (see \cite{Lahoche:2015ola}-\cite{Lionni:2016ush}). The intermediate field is a matrix-like field, and the colored edges of our diagrams are effective edges of these fields. An important consequence is that it reveals another limitation of the sector of melons that we study: it corresponds to an effective matrix-like model.

\section{Improved local truncation versus ordinary truncation}\label{section6}

Until now, the truncation around local interactions was the only way to extract non-perturbative information from the Wetterich equation in the TGFT context. The computation of the flow equations for the $\phi^4$ truncation and  for $d=5$ has been scrutinized in appendix \ref{AppB}. For these computation we used of the standard modified Litim's regulator :
\begin{equation}
r_s(\vec{p}\,)=Z(s) (e^{2s}-\vec{p}\,^2)\Theta(e^{2s}-\vec{p}\,^2)\,,
\end{equation}
where $\Theta$ is the Heaviside step function. Other standard choices could have been considered, and have been investigated in literature \cite{Litim:2000ci}. For our purpose we only consider this choice, which can be viewed as a first limitation of our result. Indeed, even if the full renormalization group flow must be independent of the choice of the regulator, it is well known that approximations used to solve the exact renormalization group equation introduce a spurious dependence on the regulator. \\

\noindent
In the UV limit, the flow equation for the local $\pi^4$ truncation are  ($\Omega_{d-1}=\pi^2/2$ for $d=5$):
\begin{equation}\label{melotruncation}
\left\{
    \begin{array}{ll}
       \beta_m &= -(2+\eta_s)\bar{m}^2-\frac{10\bar\lambda\pi^2}{(1+\bar{m}^2)^2}\Big(1+\frac{\eta_s}{6}\Big) \\
        \beta_\lambda &=-2\eta_s\bar{\lambda}+\frac{4\bar\lambda^2\pi^2}{(1+\bar{m}^2)^3}\Big(\frac{\eta_s}{6}+1\Big)\,.
    \end{array}
\right.
\end{equation}
where the anomalous dimension is given by:
\begin{equation}
 \eta_s=\frac{4\bar\lambda\pi^2}{(1+\bar{m}^2)^2-\bar{\lambda}\pi^2}\,.
\end{equation}
Truncation proceed  with a systematic projection into a finite dimensional region of the full theory space. For the $\phi^4$ truncation, we combine both a derivative expansion and a mean field expansion, and therefore the  average effective action is assumed to be of  the form:
\begin{equation}
\Gamma_s=Z(s)\sum_{\vec p} T_{\vec p}(\vec p\,^2+e^{2s}\bar{m}^2(s))\bar{T}_{\vec p}+Z^2(s)\bar{\lambda}(s)\sum_{i}\mathcal{V}^{(i)}[T,\bar{T}]
\end{equation}
where the $\mathcal{V}^{(i)}[T,\bar{T}]$ denotes the quartic melonic interactions:
\begin{equation}
\mathcal{V}^{(i)}[T,\bar{T}]=\sum_{\{\vec{p}_k\,,k=1,\cdots,4\}} \delta_{p_{1i}p_{4i}}\delta_{p_{2i}p_{3i}}\left(\prod_{j\neq i}\delta_{p_{1j}p_{2j}}\delta_{p_{4j}p_{2j}}\right) T_{\vec{p}_1}\bar{T}_{\vec{p}_2}T_{\vec{p}_3}\bar{T}_{\vec{p}_4}\,.
\end{equation}
The flow equation \eqref{melotruncation} are obtained from the Wetterich-Morris equation from an expansion of the right hand side in powers of the fields, and after systematic identification we only retains the melonic diagrams i.e. the relevant contributions in the UV limit. Hence, one can think that we recover exactly the same information by
 using the melonic structure equations.
Note also that  the melonic structure being incremented from the melonic diagrams and  keeping to build the renormalization group flow. The flow equations are computed using a local approximation. Then, the momentum dependence of the 1PI $n$-points functions is lost. This dependence play an important role in the computation of the anomalous dimension, through the term $\partial\pi^{(1)}_{00}/\partial p_1^2$. More precisely, we may writes the following corollary of proposition \ref{flowstructure}:
 We deduce the result about $\phi^4$ truncations:
\begin{corollary}
The $\pi^4$ local truncation remains a good approximation as long as:
\begin{equation}
\bar{\lambda} \left[\frac{\bar{\mathcal S}_{1,2}\bar{\mathcal{A}}_{3s}}{\bar{\mathcal S}_{1,3}}\right]\ll1\,, \mbox{and}\qquad \bar{\lambda} \left\vert\frac{\bar{\mathcal{S}}_{0,2}^\prime}{ \bar{\mathcal{S}}_{1,2}^{\prime} }\right\vert \ll1.
\end{equation}
\end{corollary}

\noindent
We will estimate the results of this section by using  two methods, and discuss the corrections coming from the improvements of the structure equations. Let us start  with a summary of the results provided from the $\phi^4$ truncation. Note that the terminology $\phi^4$ truncation is abusive to talk about the improved flow equation. Indeed, roughly speaking the flow is not truncated in the subspace of the local melonic interactions generated from the $\phi^4$ melonic interactions following the receipt given from definition \eqref{figmelons} and proposition \eqref{cormelons}. The flow of all these melonic couplings is entirely controlled from the flow of the $\phi^4$ melonic interaction. Then, keeping in mind that  in the subspace of the theory space that we consider, the choice of the initial conditions and the restriction to the UV sector is generated by the only approximations that we make on  $\Gamma_s^{(2)}(\vec{p}\,)$. The closed equation given in the  proposition \eqref{lemmatwo} cannot be solved directly, and the approximation  that we propose here consist to make a choice for this function. In the symmetric phase, the derivative expansion seems to be  natural, and the flow of each parameters that we introduced can be computed following a straightforward generalization of the proof of proposition \ref{flowstructure}. A simple choice, close to the one of the $\phi^4$ truncation may be:
\begin{equation}
\Gamma_s^{(2)}(\vec{p}\,)=Z(s)(\vec{p}\,^2+e^{2s}\bar{m}^2)\,.
\end{equation}
This choice neglects all the contributions of order $\mathcal{O}(\vec{p}\,^2)$ in the derivative expansion. The standard proof involve renormalizability of the model : In the sector $\Lambda\ll k\ll1$, we expect that only renormalizables interactions survive, i.e.  all operators with positive or null canonical dimension. 

\subsection{Ordinary $\phi^4$ truncation}
 The system \eqref{melotruncation} has a trivial fixed point: (the Gaussian fixed point (gfp)). Expanded the beta function around this point, we can  show that the flow is asymptotically free, i.e., the beta function $\beta_\lambda$ is negative:
\begin{equation}
\beta_\lambda \approx -4\pi^2\bar{\lambda}^2+\mathcal{O}(\bar{\lambda}^2, \bar{m}^2)\,.
\end{equation}
In addition to the Gaussian fixed point, we find two non-Gaussian fixed points, say $ng1$ and $ng2$ ($ngi=(\bar{m}^2_i,\bar{\lambda}_i)$, $i=1,2$):
\begin{equation}
ng1\approx (-0.52,0.0028)\,,\qquad ng2\approx (-0.87,0.0036)\,.
\end{equation}
Let $\beta_i=(\beta_m,\beta_\lambda)$ and $u_i=(\bar{m}^2,\bar{\lambda})$. The stability matrix $\beta_{ij}:=\partial \beta_j/\partial u_i$ may be computed around each fixed point, as its eigenvalues and eigenvectors. The opposite values of the eigenvalues are \textit{critical exponents}, and we find two eigenvalues  all times, say $\theta^+$ and $\theta_-$:
\begin{equation}
(\theta^+,\theta^-)\vert_{gfp}=(2,0)\,,\quad (\theta^+,\theta^-)\vert_{ng1}\approx(-4.9, 0.9)\,,\quad (\theta^+,\theta^-)\vert_{ng2}\approx(232.3, 6.9)\,.
\end{equation}
In addition, we have the values of the anomalous dimensions:
\begin{equation}
\eta_{fp1} \approx 0.55\,,\qquad \eta_{fp2}\approx -7,22\,.
\end{equation}
Then, the Gaussian fixed point has one relevant and one marginal direction, the first non-Gaussian fixed point has one relevant and one irrelevant direction, and finally the second non-Gaussian fixed point has two irrelevant directions. The irrelevant direction around $fp1$ span a critical line, separating two different phases and reaching the Gaussian fixed point such that this phenomena is  reminiscent with respect to  the well-known Wilson-Fisher fixed point. The last non-Gaussian fixed point is localized under the singularity line whose equation is given by the denominator of $\eta_s$: $(1+\bar{m}^2)^2-\bar{\lambda}\pi^2=0$. All the points in this region are disconnected from the Gaussian fixed point, which is in the symmetric phase. Then the truncated flow does not reach any point in the region under this singularity line, starting in the region upper this line. The figure \ref{figflow1} summarize these properties. For an extended discussion, see \cite{Benedetti:2015yaa} and references therein.  

\begin{center}
\includegraphics[scale=0.6]{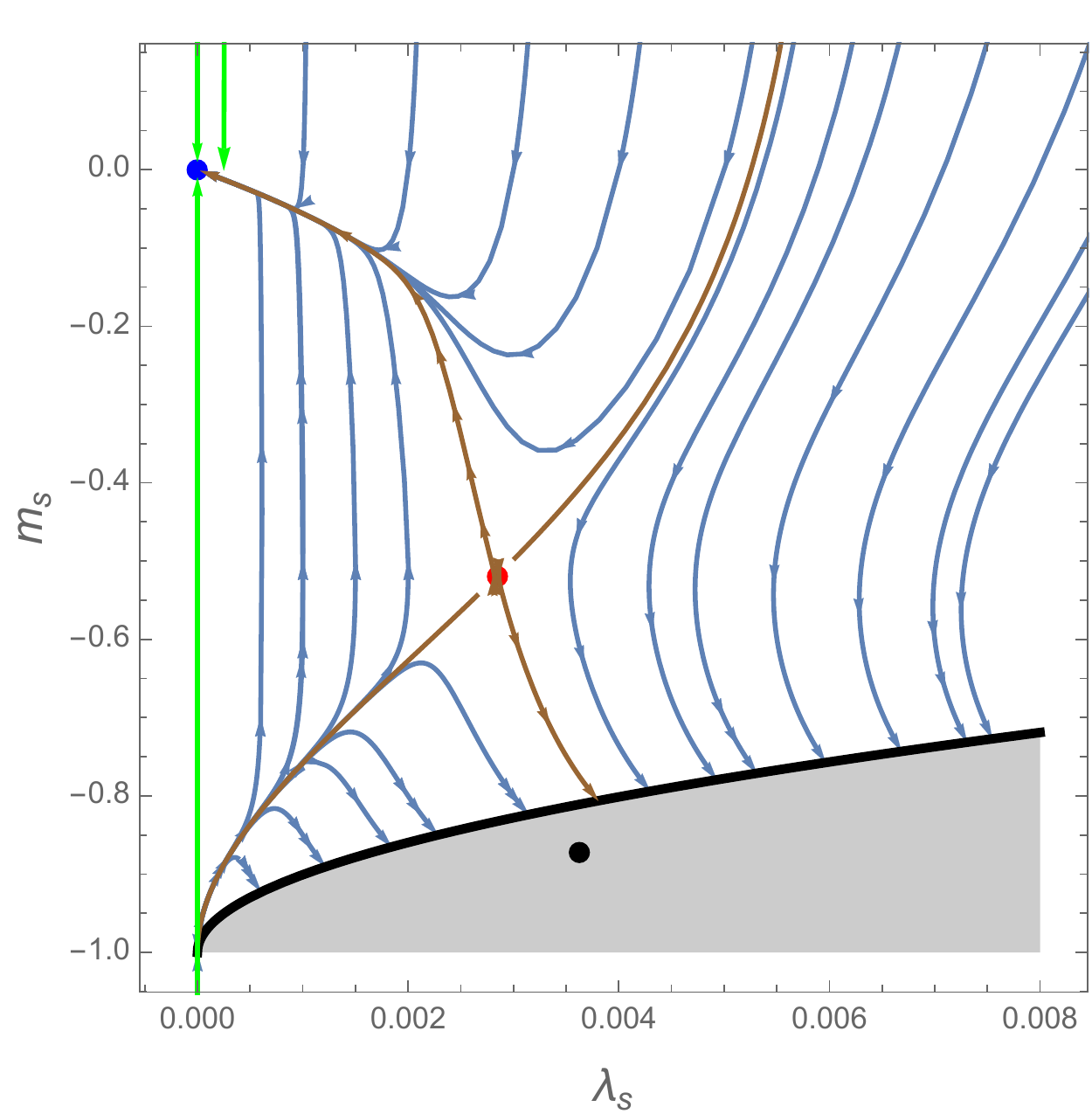} 
\captionof{figure}{Renormalization group flow trajectories around the relevant fixed points obtained from a numerical integration. The Gaussian fixed point and the first non-Gaussian fixed point are respectively in blue and in red, and  the last fixed point is in black. This fixed point is in the grey region bounded by the singularity line corresponding to the denominator of $\eta_s$. Finally, in green and brown we draw the eigendirections around Gaussian and non-Gaussian fixed points respectively. Note that arrows this fixed point  the flow are oriented from IR to UV.}\label{figflow1}
\end{center}

\subsection{Improved $\phi^4$ truncation}

We will now investigate the discussions to get  some improvement of the  truncation. Note that Litim's regulator simplifies the computation of the sums given in the previous subsection.  For instance, using the same integral approximation as in Appendix \ref{AppB}, $\bar{\mathcal{S}}_{1,2}$ and $\bar{\mathcal{S}
}_{1,3}$ are given by:
\begin{equation}
\bar{\mathcal{S}}_{1,2}=\frac{\pi^2}{(1+\bar{m}^2)^2}\left(\frac{\eta_s}{d+1}+1\right)\,,\qquad \bar{\mathcal{S}}_{1,3}=\frac{\pi^2}{(1+\bar{m}^2)^3}\left(\frac{\eta_s}{d+1}+1\right)\,.
\end{equation}
The sum $\bar{\mathcal{S}}_{1,2}^\prime$ has been also computed. The sums $\bar{\mathcal{A}}_{3s}$ and the derivative $\mathcal{S}_{0,2}^\prime$ can be given easly. First of all  $\bar{\mathcal{A}}_{3s}$ is convergent and is  given by
\begin{equation}
\bar{\mathcal{A}}_{3s}=\frac{1}{2}\frac{\pi^2}{1+\bar{m}^2}\left[\frac{1}{(1+\bar{m}^2)^2}+\left(1+\frac{1}{1+\bar{m}^2}\right)\right]\,.\label{A3comput}
\end{equation}
To compute the last sum, i.e. For the computation of the last sum $\mathcal{S}_{0,2}^\prime$, we separates the terms depending on the cut-off as :
\begin{equation}
\mathcal{A}_s(p_1)=\sum_{\vec{p}_\bot} \left[\frac{\Theta(k^2-\vec{p}\,^2)}{(Zk^2+m^2)^2}+\frac{\Theta(\vec{p}\,^2-k^2)}{(Z\vec{p}\,^2+m^2)^2}\right]\,.
\end{equation}
The last term in the above relation is divergent, and then requires an UV regularization. Computing the derivative with respect to $p_1^2$, the regulator dependent terms are canceled in the deep UV, and we get:
\begin{eqnarray*}
\frac{d\mathcal{A}_s}{dp_1^2}(p_1=0):&=&\Omega_{d-1}\frac{d}{dp_1^2}\left[\frac{(k^2-p_1^2)^{\frac{d-1}{2}}}{(Zk^2+m^2)^2}+(d-1)\int_{\sqrt{k^2-p_1^2}}^{+\infty}\frac{x^{d-2} dx}{(Zx^2+m^2)^2}\right]\bigg\vert_{p_1=0}\cr
&+&(d-1)\Omega_{d-1}\int_k^\Lambda x^{d-1}dx\frac{d}{dp_1}\Big[\frac{1}{(Z(x^2+p_1^2)+\bar m^2)^2}\Big]\Big|_{p_1=0,\Lambda\rightarrow \infty}\,.
\end{eqnarray*}
Setting $d=5$, we get
\begin{equation}
\frac{d\mathcal{A}_s}{dp_1^2}(p_1=0)=-\frac{\pi^2}{2k^2 Z^2}\frac{1}{(1+\bar{m}^2)}\Big(1+\frac{1}{1+\bar m^2}\Big)\,.
\end{equation}
By adding  these sums in the improved flow equations, we get  the beta functions:
\begin{equation}
\left\{
    \begin{array}{ll}
       \beta_m &= -(2+\eta_s)\bar{m}^2-10\bar{\lambda}\frac{\pi^2}{(1+\bar{m}^2)^2}\left(\frac{\eta_s}{6}+1\right) \\
        \beta_\lambda &=-2\eta_s\bar{\lambda}+4\bar{\lambda}^2\frac{\pi^2}{(1+\bar{m}^2)^3}\left(\frac{\eta_s}{6}+1\right)\left[1-\bar{\lambda} \pi^2\left(\frac{1}{(1+\bar{m}^2)^2}+\left(1+\frac{1}{1+\bar{m}^2}\right)\right)\right]\,.\label{betaimproved}
    \end{array}
\right. 
\end{equation}
The anomalous dimension may be computed in the same way using the result of proposition \ref{flowstructure}. we get simply
\begin{equation}
\eta_s=4\bar{\lambda}\pi^2\,\frac{(1+\bar m^2)^2-\frac{1}{2}\bar{\lambda}\pi^2(2+\bar m^2)}{(1+\bar{m}^2)^4+\frac{1}{3}\bar\lambda^2\pi^4(2+\bar m^2)-\pi^2\bar{\lambda}(1+\bar m^2)^2}\,.\label{etaimproved}
\end{equation}
The flow equations \eqref{betaimproved} and \eqref{etaimproved} may be investigated numerically as their truncated counterpart \eqref{melotruncation}.  From now we can  recover the singularity of the line, and its flow equation is exactly the same. This result  comes  from the symmetric phase but  not for  the restriction on the number of interactions. In the second part, we find a fixed point with one relevant and one irrelevant direction, say $ng1^\prime$, exactly as for the crude truncation. Moreover, the coordinates, as well as the critical exponents in the anomalous dimension remains very close to their corresponding values for truncation:
\begin{equation}
ng1^\prime\approx(-0.55, 0.003)\,,\qquad \eta_{ng1^\prime}\approx 0,63\,,\qquad (\theta^+,\theta^-)\vert_{ng1^\prime}\approx (-3.91, 0.86)\,.
\end{equation}
Finally, the third fixed point $ng2$ which was  localized under the singularity line has been completely discarded, and we can  explan it as a consequence of the crude truncation.  The   figure \ref{figflow1} -- on right is very similar to what we obtain using the crude truncation. In particular, we strengthen the conclusions about the existence of a non-trivial fixed point in the phase-space, which behaves like a Wilson-Fisher fixed point. The occurrence of such a fixed point has been considered as an important feature because it advocate a phase transition, which plays an important role in the space-time emergence following the \textit{geometrogenesis scenario} (see introduction or \cite{Oriti:2013jga}-\cite{Wilkinson:2015fja} for more details). \\

\noindent
Let us remark that, if  a fixed point is stable by adding the higher order interactions, this fixed point can  be taking into account and is called a true fixed point. But we see from proposition \ref{prophigher} that any fixed point for $\bar{m}^2$ and $\bar{\lambda}$ is a fixed point for renormalized higher interactions. Indeed, consider for instance $\bar{\pi}_3$ given by expression \eqref{exppi3}:
\begin{equation}
\bar{\pi}_3=16\bar{\lambda}^3(s)\bar{\mathcal{A}}_{3s}\,.
\end{equation}
 We have seen explicitly using expression \eqref{A3comput} that $\bar{\mathcal{A}}_{3s}$ only depends on $\bar{m}^2$, then:
\begin{equation}
\beta_\lambda=\beta_m=0\,\to\, \dot{\bar{\pi}}_3=0\,.
\end{equation}
The same argument holds for higher interactions: all the $\bar{\mathcal{A}}_{Ns}$ only depends on $\bar{m}^2$, so that any fixed point for mass and quartic melonic coupling is a true fixed point for all the local interactions built from them. To put in a nutshell:
\begin{claim}
In the UV sector, and at the first non-trivial order in the derivative expansion, there exist a critical non-Gaussian fixed point in the subspace of local interactions generated from the quartic melonic  ones.
\end{claim}

\section{Discussions and conclusion}\label{section7}
 In this paper we have studied the Wetterich flow equation for the TGFT models. In particular we consider the $T_5^4$ model defined without gauge projection. This model is showed to be just renormalizable. Using the symmetric properties of the model, the first, second WT-identities are derived. On the other hand the Wetterich flow equation is discussed. In the symmetric phase and in the melonic sector  the WT-identities are used to derive the so call structure equations. As physical consequence, these equations can help to improve the truncation for this TGFT model. The conjecture that the N-WT-identities maybe derived from the first-WT-identity by derivative maybe used to show if or not the fixed point is stable. For instance in the case of $T_5^4$ TGFT model we have showed that the fixed point is stable. Note also that this is the case where the higher order interaction is added to the the theory. Also, the flow equations can be translated in the autonomous system of differential equation which leads to the numerical computation of other fixed point. The trajectory behavior around these fixed point is also discussed.

As a discussion,  first of all, we only considered the symmetric phase, and the influence of $\Gamma^{(n,m)}_s\,\,,n\neq m$, which are of order $M$ and  is completely discarded from our analysis. Moreover $M\neq 0$ maybe introduce a non-trivial dependence for the wave function renormalization: $Z\to Z(M,\bar{M})$. On the other hand   we only consider the lowest terms in the derivative expansion for $\Gamma^{(2)}_s$. A first way of investigation could be then to explore the robustness of our conclusion for higher truncation in the derivative expansion.  We only consider the local potential approximation, i.e. the potential which can be expanded as an infinite sum of connected melons generated from quartic melonic interaction. But the power counting show that, by chosing the truncation at any order over the melonic subspace, the terms coming from the derivative expansion  in a  local interactions will contribute of the same footing as  higher local interactions. Then,  we have to consider the full momentum dependence of the 1PI functions.  Finally, we restrict our attention to the melonic sector, which is know to be the relevant sector in the UV. However, as for the deviations from ultralocal interactions, for a given truncation, non melonic bubbles have to be treated on the same way as higher melonic interactions with respect to the power counting (i.e. they could have the same canonical dimension). For these reasons, the conclusion of this paper  becomes partial  and we will investigate in  forthcoming work the discussion about that.

\section*{Acknowledgments}
D Ousmane Samary research   is supported by the Alexander von Humboldt foundation.
 
\appendix
\begin{center}
\begin{Large}
\textbf{Appendix}
\end{Large}
\end{center}
\section{Redundancy of the Ward-Takahashi identities}\label{AppA}

In this section we  discuss the WT-identity for zero-momenta $6$-point function.  This equation  have the same information than the melonic structure equation for $\pi_3$, given by Lemma \ref{lemmasix} or equation \eqref{exppi3}. One can prove  that this similarities  can be  generalized in the melonic sector, that we summarize in the following  statement:
{\it In the melonic sector, the structure equations for $\pi_N$ contains no more information than the WT-identities.} 

For $\pi_3$, this result can be show directly. In the first time, we have to derive the WT-identity for $\phi^6$ interaction. Our analysis in the next appendix requires the  \textit{second} WT-identity, obtained from the WT-identity applying the fourth derivative $\frac{\partial^{4}}{\partial M_{\vec p_1}\partial \bar M_{\vec p_2}\partial M_{\vec p_3}\partial \bar M_{\vec p_4}}$. From the same strategy as for the first WI-identity, we get:
\bea
&&\sum_{\vec p_\bot,\vec p\,'_\bot}\delta_{\vec p_\bot\vec p\,'_\bot}\Delta C_s(\vec p,\vec p\,')\Big[\Gamma^{(6)}_{s,\vec p_4\vec p_2\vec p\,';\vec p_3\vec p_1\vec p}-4\Gamma^{(4)}_{s,\vec p\vec p_4;\vec p_1\vec p\,''}G_s(\vec p\,'')\Gamma^{(4)}_{s,\vec p\,''\vec p_2;\vec p_2\vec p\,'}\Big]G_s(\vec p)G_s(\vec p\,')\cr
&&+\sum_{\vec p_\bot,\vec p\,'_\bot}\delta_{\vec p_\bot\vec p\,'_\bot}\Big[-\delta_{\vec p_3 \vec p\,'}\Gamma^{(4)}_{s,\vec p_2\vec p_4;\vec p\vec p_1}
-\delta_{\vec p_1\vec p\,'}\Gamma^{(4)}_{s,\vec p_2\vec p_4;\vec p\vec p_3}
+\delta_{\vec p_4 \vec p}\Gamma^{(4)}_{s,\vec p\,'\vec p_2;\vec p_1\vec p_3}
+\delta_{\vec p_2 \vec p}\Gamma^{(4)}_{s,\vec p\,'\vec p_4;\vec p_1\vec p_3}\Big]=0\,.\cr
&&\label{sixward}
\eea
The \textit{second WT-identity} then gives a relation between $6$ and $4$ points functions. As for the first WT-identity, we will interested by the zero-momenta version. To this end, we introduce  the quantity  $X_{\vec p_1 \vec p_2\vec p_3 \vec p_4}$ defined as:
\bea
X_{\vec p_1 \vec p_2\vec p_3 \vec p_4}&=&-\delta_{\vec p_3 \vec p\,'}\Gamma^{(4)}_{s,\vec p_2\vec p_4;\vec p\vec p_1}
-\delta_{\vec p_1\vec p\,'}\Gamma^{(4)}_{s,\vec p_2\vec p_4;\vec p\vec p_3}
+\delta_{\vec p_4 \vec p}\Gamma^{(4)}_{s,\vec p\,'\vec p_2;\vec p_1\vec p_3}
+\delta_{\vec p_2 \vec p}\Gamma^{(4)}_{s,\vec p\,'\vec p_4;\vec p_1\vec p_3}\cr
&=&\sum_i\Bigg\{-\pi^{(i)}_{p_2^i p^i}\Big[\delta_{\vec p_3\vec p\,'}\sym W^{(i)}_{\vec p_2\vec p_4;\vec p\vec p_1}+\delta_{\vec p_1\vec p\,'}\sym W^{(i)}_{\vec p_2\vec p_4;\vec p\vec p_3}\Big]\cr
&&-\pi^{(i)}_{p'^i p_1^i}\Big[\delta_{\vec p_4\vec p}\sym W^{(i)}_{\vec p\,'\vec p_2;\vec p_1\vec p_3}+\delta_{\vec p_2\vec p}\sym W^{(i)}_{\vec p\,'\vec p_4;\vec p_1\vec p_3}\Big]\Bigg\}\,.
\eea
Setting  $\vec p_3=\vec p_4=\vec 0$ yield to
\bea
X_{\vec p_1 \vec p_2\vec 0 \vec 0}=\sum_i\delta_{\vec p_\bot\vec p\,'_\bot}\Bigg\{-\pi^{(i)}_{p_2^ip\,'^i}\Big[\delta_{\vec p\,'\vec 0}\sym W^{(i)}_{\vec p_2\vec 0;\vec p\vec p_1}+\delta_{\vec p_1\vec p\,'}\sym W^{(i)}_{\vec p_2\vec 0;\vec p\vec 0}\Big]\cr
-\pi^{(i)}_{p\,'^ip_1^i}\Big[\delta_{\vec 0\vec p}\sym W^{(i)}_{\vec p\,'\vec p_2;\vec p_1\vec 0}+\delta_{\vec p_2\vec p}\sym W^{(i)}_{\vec p\,'\vec 0;\vec p\vec 0}\Big]\Bigg\}
\eea
Now, setting $\vec p_1=(p_1',\vec 0_\bot)$, $\vec p_2=(p_2,\vec 0_\bot)$, we get:
\bea
X_{\vec p_1 \vec p_2\vec 0 \vec 0}&=&\sum_{i}\delta_{\vec p_\bot\vec p\,'_\bot}\Big\{-\pi^{(i)}_{p_2^ip^i}\Big[\delta_{\vec p\,'\vec 0}\delta_{\vec p_{2\bot i}\vec p_{\bot i}}\delta_{\vec 0_{\bot i}\vec p_{1\bot i}}\delta_{\vec p_{2 i}\vec p_{1 i}}\delta_{\vec 0\vec p_{ i}}+\delta_{\vec p\,'\vec 0}\delta_{\vec 0_{\bot i}\vec p_{\bot i}}\delta_{\vec p_{2\bot i}\vec p_{1\bot i}}\delta_{\vec 0\vec p_{1 i}}\delta_{\vec p_{2i}\vec p_{ i}}\cr
&&\delta_{\vec p\,'\vec p_1}\delta_{\vec p_{2\bot i}\vec p\,'_{\bot i}}\delta_{\vec 0 \vec p_{2}}\delta_{\vec 0\vec p\,'_{ i}}+\delta_{\vec p\,'\vec p_{1}}\delta_{\vec p_{2\bot }\vec 0_{\bot }}\delta_{\vec p_{\bot }\vec 0_{\bot }}\delta_{\vec p_{2 i}\vec p_{ i}}\Big]\cr
&&-\pi^{(i)}_{p_1^ip\,'^i}\Big[\delta_{\vec p\vec 0}\delta_{\vec p\,'_{\bot i}\vec p_{1\bot i}}\delta_{\vec p_{2\bot i}\vec 0_{\bot i}}\delta_{\vec p\,'_{ i}\vec 0}\delta_{\vec p_{2i}\vec p_{ 1i}}+\delta_{\vec p\vec 0}\delta_{\vec p_{2\bot i}\vec p_{1\bot i}}\delta_{\vec p\,'_{\bot i}\vec 0_{\bot i}}\delta_{\vec p\,'_i\vec p_{1 i}}\delta_{\vec p_{2i}\vec 0}\cr
&&\delta_{\vec p\vec p_2}\delta_{\vec p\,'_{\bot i}\vec 0_{\bot i}}\delta_{\vec p_{1\bot i} \vec 0_{\bot i}}\delta_{\vec p\,'_i\vec p_{1 i}}+\delta_{\vec p\vec p_{2}}\delta_{\vec p\,'_{\bot i }\vec p_{1\bot i}}\delta_{\vec p\,'_{i }\vec 0 }\delta_{\vec p_{1 i}\vec 0}\Big]\Big\}\,.
\eea
By adding these two result in the equation \ref{sixward}, we deduce:

\begin{corollary} \textbf{Second zero momenta Ward-Takahashi identity:}\label{secondWI}
In the symmetric phase, the zero-momenta $6$-point fonction satisfies:
\bea
\sum_{\vec p_\bot}Z_{-\infty} \bigg(1+\frac{\partial \tilde{r}_{s}(\vec{p}_{\bot})}{\partial p_1^2}\bigg)G^2_s(\vec{p}_\bot)\Big[\frac{1}{3}\Gamma^{(6)\,,1}_{s,\vec 0\vec 0\vec 0;\vec 0\vec 0\vec 0}-(\Gamma^{4,1}_{s,\vec 0\vec 0;\vec 0\vec 0})^2G_s(\vec{p}_\bot)\Big]=-2\frac{d}{dp_1^2}\pi_{sp_1p_1}^{(1)}\vert_{p_1=0}\,.
\eea
where $\Gamma^{(3,3)\,,i}_{s,\vec 0\vec 0\vec 0;\vec 0\vec 0\vec 0\vec 0}$ denote the component of the melonic $6$-point function which build the graphs with heart external faces of color $i$. 
\end{corollary}

\noindent
Note that the two factors $1/3$  comes from the fact that we only keep the leading order contributions. For instance, if we consider the $6$-point function, there are $(3!)^2$ allowing the permutation of the external variables. However, the leading order contractions are such that $\vec{p}$ and $\vec{p}\,^\prime$ are on the same vertex, and  this leads to  $3$ possibilities, times the number of configurations of the remaining variables i.e.  $2$. Hence, there are only $3!$ melonic contraction among the $(3!)^2$ allowed.\\

Because $\Gamma^{(6),1}_{s;\vec{0}\vec{0}\vec{0},\vec{0}\vec{0}\vec{0}}=\pi_3\times 3!$ and $\Gamma^{(4),1}_{s;\vec{0}\vec{0},\vec{0}\vec{0}}=4\lambda(s)$, the WI-identity maybe rewritten as
\begin{equation}
Z_{-\infty}\left(\mathcal{L}_s\pi_3-8\lambda^2(s)\,\mathcal{U}_s\right)=-\frac{d}{dp_1^2}\pi_{sp_1p_1}^{(1)}\vert_{p_1=0}\,. 
\end{equation}
The right-hand-side may be computed directly from the definition of $\pi_{sp_1p_1}^{(1)}$
\begin{equation}
\frac{d}{dp_1^2}\pi_{sp_1p_1}^{(1)}\vert_{p_1=0}=-4\lambda^2(s) \frac{d}{dp_1^2}\mathcal{A}_s(p_1=0)\,,
\end{equation}
such that
\begin{equation}
Z_{-\infty}\mathcal{L}_s\pi_3=4\lambda^2(s)\bigg[2Z_{-\infty}\mathcal{U}_s+\frac{d}{dp_1^2}\mathcal{A}_s(p_1=0)\bigg]\,. \label{wiwi}
\end{equation}
Because $\pi_3$ is at least of order $3$ in $\lambda$, the last term have to be of order $1$. To be more precise let us introduce the following definition   of the quantity $\mathcal{U}_s$, 
\begin{align}
\nonumber  2 Z_{-\infty}\mathcal{U}_s+\frac{d}{dp_1^2}\mathcal{A}_s(p_1=0)&=\sum_{\vec{p}_\bot}\left[2\left(Z_{-\infty}+\frac{dr_s}{dp_1^2}(p_1=0)\right)-2\left(Z(s)+\frac{dr_s}{dp_1^2}(p_1=0)\right)\right]G^3_s(\vec{p}_\bot) \,,
\\\nonumber&=2\sum_{\vec{p}_\bot}\left(Z_{-\infty}-Z(s)\right)G^3_s(\vec{p}_\bot)\,,
\\&=4Z_{-\infty}\lambda(s)\mathcal{L}_s \sum_{\vec{p}_\bot}G^3_s(\vec{p}_\bot)\,,
\end{align}
where  we used the relation $Z(s)=Z_{-\infty}(1-2\lambda(s)\mathcal{L}_s)$. Then the WT-identity \eqref{wiwi} is written  as:
\begin{equation}
Z_{-\infty}\mathcal{L}_s\pi_3=Z_{-\infty}\mathcal{L}_s\times 16\lambda^3(s) \sum_{\vec{p}_\bot}G^3_s(\vec{p}_\bot)\,.
\end{equation}
Therefore we recover the structure equation for $\pi_3$ givin in  \eqref{exppi3} by  simplifying the  factor $Z_{-\infty}\mathcal{L}_s$.

\section{$\phi^4$ truncation for $T^4_5$-U(1) model}\label{AppB}
In this section we  derive the flow equation for $\phi^4$ melonic truncation. The procedure is now standard in the TGFT literature, and we only indicate the main steps, the references \cite{Geloun:2016qyb}-\cite{Lahoche:2016xiq} maybe consulted for more details. The  truncation is a systematic projection of the renormalization group flow into a finite dimensional subspace of the full theory space. \\

\noindent
For $d=5$ we consider the $T^4_5$-U(1) truncation given for the effective action  ($k=e^s$) by:
\bea
\Gamma_s=Z(s)\sum_{\vec p} T_{\vec p}(\vec p\,^2+e^{2s}\bar{m}^2(s))\bar{T}_{\vec p}+Z^2(s)\bar{\lambda}(s)\sum_{i}\mathcal{V}^{(i)}[T,\bar{T}]\,,
\eea
where we used renormalized couplings $Z^2(s)\bar{\lambda}(s)$, and $\sum_i\mathcal{V}^{(i)}[T,\bar{T}]=:V[T,\bar T]$ is the quartic melonic potential. The Wetterich equation \eqref{wetterich} can be formally expand as:
\bea
\partial_s\Gamma_s=\Tr\,\dot r_s G^{(0)}_s\left[1-2\lambda(s)V''[T,\bar T]G^{(0)}_s+4\lambda^2(s)V''[T,\bar T]G^{(0)}_sV''[T,\bar T]G^{(0)}_s+\cdots\right]\,.
\eea
The flow equations involves many contractions of lines, and maybe pictured as sums of the following diagrams
\begin{equation}
\partial_s\Gamma^{(2)}_s=\sum_{i=1}^d\left\{K_1^{(2)}\,\,\vcenter{\hbox{\includegraphics[scale=0.6]{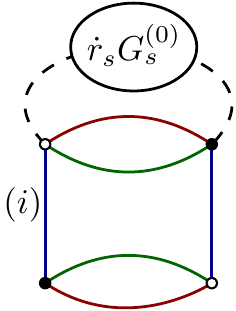} }}+K_2^{(2)}\,\,\vcenter{\hbox{\includegraphics[scale=0.6]{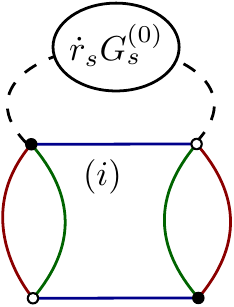} }}\right\}\,,\label{1}
\end{equation}
\begin{equation}
\partial_s\Gamma^{(4)}_s=\sum_{i,j}\left\{K_1^{(4)ij}\vcenter{\hbox{\includegraphics[scale=0.6]{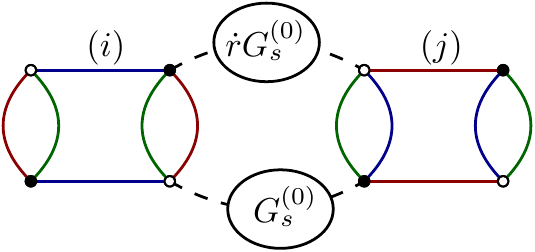}  }}+K_2^{(4)ij}\vcenter{\hbox{\includegraphics[scale=0.6]{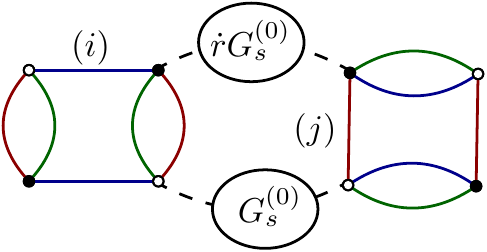}  }}+K_3^{(4)ij}\vcenter{\hbox{\includegraphics[scale=0.6]{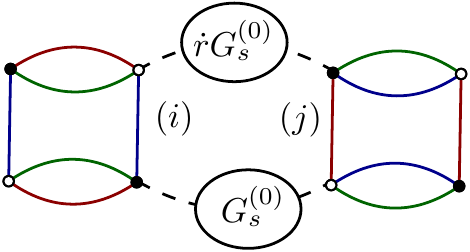}  }}\right\}\,,\label{2}
\end{equation}
where $K^{(2)}_1, K_2^{(2)},...,K_3^{(4)ij}$ are numerical coefficients. In the deep UV sector, we have retained only the melonic contributions. They comes from the first term on the right hand side of equation \eqref{1}, and from the first term of the right hand side for equation \eqref{2}, with the condition that  $i=j$. Moreover, because of the choice of the truncation, we have the following relations:
\bea
m^2(s)=\Gamma_{s}^{(2)}(\vec 0),\quad Z(s)=\frac{d\Gamma_{s}^{(2)}(\vec 0)}{dp_1^2},\quad \partial_s \Gamma^{(4),1}_{s,\vec 0\vec 0;\vec 0\vec 0}=4 \partial_s\lambda(s)\,,
\eea
leading to the following
\bea
&&\partial_sm^2(s)=-2d\lambda(s) \sum_{\vec p_\bot}\frac{\dot r_s(\vec p_\bot)}{[Z\vec p\,^2_\bot+m^2+r_s(\vec p_\bot)]^2}\\
&&\partial_s Z(s)=-2\lambda(s)\frac{d}{dp_1^2} \sum_{\vec p_\bot}\frac{\dot r_s(\vec p)}{[Z\vec p\,^2+m^2+r_s(\vec p\,)]^2}\Big|_{p_1=0}\cr
&&4\partial_s\lambda(s)=16\lambda^2(s)\sum_{\vec p_\bot}\frac{\dot r_s(\vec p_\bot)}{[Z\vec p\,^2_\bot+m^2+r_s(\vec p_\bot)]^3}\,.
\eea
Now we have to compute the sums in the above relations. Using  the Litim modified regulator: $r_s(\vec p\,)=Z(k^2-\vec p\,^2\,)\theta(k^2-\vec p\,^2\,)$, we get:
\bea
\dot r_s(\vec p\,)=Z\Big[\eta_s(k^2-\vec p\,^2)+2k^2\Big]\theta(k^2-\vec p\,^2)\,,
\eea
such that
\bea
\partial_sm^2(s)&=&-\frac{2d\lambda(s)}{(Z k^2+m^2)^2}\sum_{\vec p_\bot}\dot r_s(\vec p_\bot)\cr
&=&-\frac{2d\lambda(s) Z}{(Z k^2+m^2)^2}\sum_{\vec p_\bot}\Big[\eta(k^2-\vec p\,^2)+2k^2\Big]\theta(k^2-\vec p\,^2)\,.
\eea
 In the continuum approximation  this sum can be given simply  for large $k$. For convenience for the rest, we introduce two sums:
\begin{equation}
S_1(p_1^2):=\sum_{\vec p_\bot}\theta(k^2-\vec p\,^2)\,,\qquad S_2(p_1^2)= \sum_{\vec p_\bot}\vec p\,^2\theta(k^2-\vec p\,^2)\,,
\end{equation}
and we get:
\bea
S_1(p_1^2)\approx (k^2-p_1^2)^{\frac{d-1}{2}}\Omega_{d-1}\,,\qquad S_2(p_1^2)\approx \left[\frac{d-1}{d+1}(k^2-p_1^2)+p_1^2\right](k^2-p_1^2)^{\frac{d-1}{2}}\Omega_{d-1}\,,
\eea
where $\Omega_{d}$ is the volume of the $d$-ball : $\Omega_{d}:=\pi^{d/2}/\Gamma(d/2+1)$. It follows:
\bea
\partial_sm_s^2=-\frac{4d\lambda Z}{(Z_s k^2+m^2)^2}\left(\frac{\eta}{d+1}+1\right)\Omega_{d-1}\,,
\eea
and the beta function $\beta_m:=\dot{\bar{m}}^2(s)$ can be straightforwardly deduced:
\bea
\beta_m=-(2+\eta_s)\bar{m}^2-\frac{4d\bar\lambda}{(1+\bar{m}^2)^2}\Big(1+\frac{\eta_s}{d+1}\Big)\Omega_{d-1}.
\eea
In the same manner we get  the flow for the wave function renormalization:
\bea
\partial_s Z(s)=-\frac{2\lambda}{(Zk^2+m^2)^2}\frac{d}{dp_1^2}\sum_{\vec p_\bot}\dot r_s(\vec p)\bigg\vert_{p_1=0}=-\frac{2\lambda}{(Zk^2+m^2)^2}\left((2+\eta_s)k^2S_1^{\prime}-\eta_sS_2^{\prime}\right)\,,
\eea
with the definitions:
\bea
S_1^{\prime}:=\frac{dS_1(p_1)}{dp_1^2}\Big|_{p_1=0}\approx -\frac{d-1}{2}k^{d-3}\Omega_{d-1},\quad S_2^{\prime}:=\frac{dS_2(p_1)}{dp_1^2}\Big|_{p_1=0}\approx -\frac{d-3}{2}k^{d-1}\Omega_{d-1}\,,
\eea
Also the anomalous dimension takes the form
\bea
\eta_s=\frac{8\bar\lambda\Omega_{d-1}}{(1+\bar{m}^2)^2-2\bar{\lambda}\Omega_{d-1}}\,.
\eea
Finally, the beta function of the coupling constant maybe computed:
\bea
\partial_s\lambda_s=\frac{16\lambda^2}{4(Zk^2+m^2)^3}\sum_{\vec p_\bot} \dot r_s(\vec p_\bot)=\frac{8\lambda^2 Z k^{d-1}\Omega_{d-1}}{(Zk^2+m^2)^3}\Big(\frac{\eta_s}{d+1}+1\Big)\,,
\eea
which leads to
\bea
\beta_\lambda=-2\eta_s\bar\lambda+\frac{8\bar\lambda^2\Omega_{d-1}}{(1+\bar{m}^2)^3}\Big(\frac{\eta_s}{d+1}+1\Big)\,.
\eea


\begin{thebibliography}{16}


\bibitem{Oriti:2006ar} 
  D.~Oriti,
  ``A Quantum field theory of simplicial geometry and the emergence of spacetime,''
  J.\ Phys.\ Conf.\ Ser.\  {\bf 67}, 012052 (2007)
  doi:10.1088/1742-6596/67/1/012052
  [hep-th/0612301].


\bibitem{deCesare:2016rsf} 
  M.~de Cesare, A.~G.~A.~Pithis and M.~Sakellariadou,
  ``Cosmological implications of interacting Group Field Theory models: cyclic Universe and accelerated expansion,''
  Phys.\ Rev.\ D {\bf 94}, no. 6, 064051 (2016)
  doi:10.1103/PhysRevD.94.064051
  [arXiv:1606.00352 [gr-qc]].



\bibitem{Gielen:2016dss} 
  S.~Gielen and L.~Sindoni,
  ``Quantum Cosmology from Group Field Theory Condensates: a Review,''
  SIGMA {\bf 12}, 082 (2016)
  doi:10.3842/SIGMA.2016.082
  [arXiv:1602.08104 [gr-qc]].



\bibitem{Gielen:2017eco} 
  S.~Gielen and D.~Oriti,
  ``Cosmological perturbations from full quantum gravity,''
  arXiv:1709.01095 [gr-qc].

\bibitem{Rovelli:2008brv} 
  C.~Rovelli,
  ``Quantum gravity,''
  Scholarpedia {\bf 3}, no. 5, 7117 (2008).
  doi:10.4249/scholarpedia.7117

\bibitem{Rovelli:2011eq} 
  C.~Rovelli,
  ``Zakopane lectures on loop gravity,''
  PoS QGQGS {\bf 2011}, 003 (2011)
  [arXiv:1102.3660 [gr-qc]].


\bibitem{Rovelli:2010bf} 
  C.~Rovelli,
  ``Loop quantum gravity: the first twenty five years,''
  Class.\ Quant.\ Grav.\  {\bf 28}, 153002 (2011)
  doi:10.1088/0264-9381/28/15/153002
  [arXiv:1012.4707 [gr-qc]].



\bibitem{Oriti:2014yla} 
  D.~Oriti, J.~P.~Ryan and J.~Thurigen,
  ``Group field theories for all loop quantum gravity,''
  New J.\ Phys.\  {\bf 17}, no. 2, 023042 (2015)
  doi:10.1088/1367-2630/17/2/023042
  [arXiv:1409.3150 [gr-qc]].


\bibitem{Gurau:2009tw} 
  R.~Gurau,
  ``Colored Group Field Theory,''
  Commun.\ Math.\ Phys.\  {\bf 304}, 69 (2011)
  doi:10.1007/s00220-011-1226-9
  [arXiv:0907.2582 [hep-th]].
 
\bibitem{Rivasseau:2016rgt} 
  V.~Rivasseau,
  ``Constructive Tensor Field Theory,''
  arXiv:1603.07312 [math-ph].


\bibitem{Rivasseau:2016zco} 
  V.~Rivasseau,
  ``Random Tensors and Quantum Gravity,''
  arXiv:1603.07278 [math-ph].



\bibitem{Rivasseau:2014ima} 
  V.~Rivasseau,
  ``The Tensor Theory Space,''
  Fortsch.\ Phys.\  {\bf 62}, 835 (2014)
  doi:10.1002/prop.201400057
  [arXiv:1407.0284 [hep-th]].

\bibitem{Rivasseau:2013uca} 
  V.~Rivasseau,
  ``The Tensor Track, III,''
  Fortsch.\ Phys.\  {\bf 62}, 81 (2014)
  doi:10.1002/prop.201300032
  [arXiv:1311.1461 [hep-th]].


\bibitem{Rivasseau:2016wvy} 
  V.~Rivasseau,
  ``The Tensor Track, IV,''
  arXiv:1604.07860 [hep-th].


\bibitem{Rivasseau:2012yp} 
  V.~Rivasseau,
  ``The Tensor Track: an Update,''
  arXiv:1209.5284 [hep-th].


\bibitem{Gurau:2011xq} 
  R.~Gurau,
  ``The complete 1/N expansion of colored tensor models in arbitrary dimension,''
  Annales Henri Poincare {\bf 13}, 399 (2012)
  doi:10.1007/s00023-011-0118-z
  [arXiv:1102.5759 [gr-qc]].



\bibitem{Gurau:2010ba} 
  R.~Gurau,
  ``The 1/N expansion of colored tensor models,''
  Annales Henri Poincare {\bf 12}, 829 (2011)
  doi:10.1007/s00023-011-0101-8
  [arXiv:1011.2726 [gr-qc]].


\bibitem{Carrozza:2012uv} 
  S.~Carrozza, D.~Oriti and V.~Rivasseau,
  ``Renormalization of Tensorial Group Field Theories: Abelian U(1) Models in Four Dimensions,''
  Commun.\ Math.\ Phys.\  {\bf 327}, 603 (2014)
  doi:10.1007/s00220-014-1954-8
  [arXiv:1207.6734 [hep-th]].


\bibitem{Carrozza:2013mna} 
  S.~Carrozza,
  ``Tensorial methods and renormalization in Group Field Theories,''
  doi:10.1007/978-3-319-05867-2
  arXiv:1310.3736 [hep-th].



\bibitem{Carrozza:2013wda} 
  S.~Carrozza, D.~Oriti and V.~Rivasseau,
  ``Renormalization of a SU(2) Tensorial Group Field Theory in Three Dimensions,''
  Commun.\ Math.\ Phys.\  {\bf 330}, 581 (2014)
  doi:10.1007/s00220-014-1928-x
  [arXiv:1303.6772 [hep-th]].


\bibitem{Geloun:2013saa} 
  J.~Ben Geloun,
  ``Renormalizable Models in Rank $d\geq 2$ Tensorial Group Field Theory,''
  Commun.\ Math.\ Phys.\  {\bf 332}, 117 (2014)
  doi:10.1007/s00220-014-2142-6
  [arXiv:1306.1201 [hep-th]].
\bibitem{Lahoche:2015tqa} 
  V.~Lahoche and D.~Oriti,
  ``Renormalization of a tensorial field theory on the homogeneous space SU(2)/U(1),''
  arXiv:1506.08393 [hep-th].


\bibitem{Lahoche:2015ola} 
  V.~Lahoche, D.~Oriti and V.~Rivasseau,
  ``Renormalization of an Abelian Tensor Group Field Theory: Solution at Leading Order,''
  JHEP {\bf 1504}, 095 (2015)
  doi:10.1007/JHEP04(2015)095
  [arXiv:1501.02086 [hep-th]].


\bibitem{Geloun:2012bz} 
  J.~Ben Geloun and E.~R.~Livine,
  ``Some classes of renormalizable tensor models,''
  J.\ Math.\ Phys.\  {\bf 54}, 082303 (2013)
  doi:10.1063/1.4818797
  [arXiv:1207.0416 [hep-th]].


\bibitem{Samary:2012bw} 
  D.~Ousmane~Samary and F.~Vignes-Tourneret,
  ``Just Renormalizable TGFT's on $U(1)^{d}$ with Gauge Invariance,''
  Commun.\ Math.\ Phys.\  {\bf 329}, 545 (2014)
  doi:10.1007/s00220-014-1930-3
  [arXiv:1211.2618 [hep-th]].


\bibitem{BenGeloun:2012pu} 
  J.~Ben Geloun and D.~Ousmane.~Samary,
  ``3D Tensor Field Theory: Renormalization and One-loop $\beta$-functions,''
  Annales Henri Poincare {\bf 14}, 1599 (2013)
  doi:10.1007/s00023-012-0225-5
  [arXiv:1201.0176 [hep-th]].

\bibitem{BenGeloun:2011rc} 
  J.~Ben Geloun and V.~Rivasseau,
  ``A Renormalizable 4-Dimensional Tensor Field Theory,''
  Commun.\ Math.\ Phys.\  {\bf 318}, 69 (2013)
  doi:10.1007/s00220-012-1549-1
  [arXiv:1111.4997 [hep-th]].



\bibitem{Geloun:2011cy} 
  J.~Ben Geloun and V.~Bonzom,
  ``Radiative corrections in the Boulatov-Ooguri tensor model: The 2-point function,''
  Int.\ J.\ Theor.\ Phys.\  {\bf 50}, 2819 (2011)
  doi:10.1007/s10773-011-0782-2
  [arXiv:1101.4294 [hep-th]].



\bibitem{Carrozza:2014rba} 
  S.~Carrozza,
  ``Discrete Renormalization Group for SU(2) Tensorial Group Field Theory,''
  Ann. Inst. Henri Poincar\'e Comb. Phys. Interact. 2 (2015), 49-112
  doi:10.4171/AIHPD/15
  [arXiv:1407.4615 [hep-th]].


\bibitem{Samary:2013xla} 
  D.~Ousmane Samary,
  ``Beta functions of  $U(1)^d$ gauge invariant just renormalizable tensor models,''
  Phys.\ Rev.\ D {\bf 88}, no. 10, 105003 (2013)
  doi:10.1103/PhysRevD.88.105003
  [arXiv:1303.7256 [hep-th]].



\bibitem{Geloun:2016qyb} 
  J.~B.~Geloun, R.~Martini and D.~Oriti,
  ``Functional Renormalisation Group analysis of Tensorial Group Field Theories on $\mathbb{R}^d$,''
  arXiv:1601.08211 [hep-th].

\bibitem{Geloun:2015qfa} 
  J.~B.~Geloun, R.~Martini and D.~Oriti,
  ``Functional Renormalization Group analysis of a Tensorial Group Field Theory on $\mathbb{R}^3$,''
  Europhys.\ Lett.\  {\bf 112}, no. 3, 31001 (2015)
  doi:10.1209/0295-5075/112/31001
  [arXiv:1508.01855 [hep-th]].




\bibitem{Benedetti:2015yaa} 
  D.~Benedetti and V.~Lahoche,
  ``Functional Renormalization Group Approach for Tensorial Group Field Theory: A Rank-6 Model with Closure Constraint,''
  arXiv:1508.06384 [hep-th].

\bibitem{Benedetti:2014qsa} 
  D.~Benedetti, J.~Ben Geloun and D.~Oriti,
  ``Functional Renormalisation Group Approach for Tensorial Group Field Theory: a Rank-3 Model,''
  JHEP {\bf 1503}, 084 (2015)
  doi:10.1007/JHEP03(2015)084
  [arXiv:1411.3180 [hep-th]].

\bibitem{Carrozza:2017vkz} 
  S.~Carrozza, V.~Lahoche and D.~Oriti,
  ``Renormalizable Group Field Theory beyond melonic diagrams: an example in rank four,''
  Phys.\ Rev.\ D {\bf 96}, no. 6, 066007 (2017)
  doi:10.1103/PhysRevD.96.066007
  [arXiv:1703.06729 [gr-qc]].



\bibitem{Carrozza:2016tih} 
  S.~Carrozza and V.~Lahoche,
  ``Asymptotic safety in three-dimensional SU(2) Group Field Theory: evidence in the local potential approximation,''
  Class.\ Quant.\ Grav.\  {\bf 34}, no. 11, 115004 (2017)
  doi:10.1088/1361-6382/aa6d90
  [arXiv:1612.02452 [hep-th]].


\bibitem{Lahoche:2016xiq} 
  V.~Lahoche and D.~Ousmane Samary,
  ``Functional renormalization group for the U(1)-T$_5^6$ tensorial group field theory with closure constraint,''
  Phys.\ Rev.\ D {\bf 95}, no. 4, 045013 (2017)
  doi:10.1103/PhysRevD.95.045013
  [arXiv:1608.00379 [hep-th]].



\bibitem{Gielen:2014uga} 
  S.~Gielen and D.~Oriti,
  ``Quantum cosmology from quantum gravity condensates: cosmological variables and lattice-refined dynamics,''
  New J.\ Phys.\  {\bf 16}, no. 12, 123004 (2014)
  doi:10.1088/1367-2630/16/12/123004
  [arXiv:1407.8167 [gr-qc]].





\bibitem{Oriti:2005tx} 
  D.~Oriti,
  ``Quantum gravity as a group field theory: A Sketch,''
  J.\ Phys.\ Conf.\ Ser.\  {\bf 33}, 271 (2006)
  doi:10.1088/1742-6596/33/1/030
  [gr-qc/0512048].



\bibitem{Wilson:1971bg} 
  K.~G.~Wilson,
  ``Renormalization group and critical phenomena. 1. Renormalization group and the Kadanoff scaling picture,''
  Phys.\ Rev.\ B {\bf 4}, 3174 (1971).
  doi:10.1103/PhysRevB.4.3174

%
\bibitem{Wilson:1971dh} 
  K.~G.~Wilson,
  ``Renormalization group and critical phenomena. 2. Phase space cell analysis of critical behavior,''
  Phys.\ Rev.\ B {\bf 4}, 3184 (1971).
  doi:10.1103/PhysRevB.4.3184

\bibitem{Wetterich:1989xg} 
  C.~Wetterich,
  ``Average Action and the Renormalization Group Equations,''
  Nucl.\ Phys.\ B {\bf 352}, 529 (1991).
  doi:10.1016/0550-3213(91)90099-J.













































%

%
%

%


\bibitem{Dona:2015tnf} 
  P.~Don\`a, A.~Eichhorn, P.~Labus and R.~Percacci,
  ``Asymptotic safety in an interacting system of gravity and scalar matter,''
  Phys.\ Rev.\ D {\bf 93}, no. 4, 044049 (2016)
  doi:10.1103/PhysRevD.93.044049
  [arXiv:1512.01589 [gr-qc]].


\bibitem{Dona:2014pla} 
  P.~Don\`a, A.~Eichhorn and R.~Percacci,
  ``Consistency of matter models with asymptotically safe quantum gravity,''
  Can.\ J.\ Phys.\  {\bf 93}, no. 9, 988 (2015)
  doi:10.1139/cjp-2014-0574
  [arXiv:1410.4411 [gr-qc]].

\bibitem{Eichhorn:2013isa} 
  A.~Eichhorn and T.~Koslowski,
  ``Continuum limit in matrix models for quantum gravity from the Functional Renormalization Group,''
  Phys.\ Rev.\ D {\bf 88}, 084016 (2013)
  doi:10.1103/PhysRevD.88.084016
  [arXiv:1309.1690 [gr-qc]].











\bibitem{BenGeloun:2011xu} 
  J.~Ben Geloun,
  ``Ward-Takahashi identities for the colored Boulatov model,''
  J.\ Phys.\ A {\bf 44}, 415402 (2011)
  doi:10.1088/1751-8113/44/41/415402
  [arXiv:1106.1847 [hep-th]].


\bibitem{Perez-Sanchez:2016zbh} 
  C.~I.~Pérez-Sánchez,
  ``The full Ward-Takahashi Identity for colored tensor models,''
  doi:10.1007/s00220-018-3103-2
  arXiv:1608.08134 [math-ph].
  
\bibitem{Samary:2014tja} 
  D.~Ousmane~Samary,
  ``Closed equations of the two-point functions for tensorial group field theory,''
  Class.\ Quant.\ Grav.\  {\bf 31}, 185005 (2014)
  doi:10.1088/0264-9381/31/18/185005
  [arXiv:1401.2096 [hep-th]].



\bibitem{Samary:2014oya} 
  D.~Ousmane Samary, C.~I.~P\'erez-S\'anchez, F.~Vignes-Tourneret and R.~Wulkenhaar,
  ``Correlation functions of a just renormalizable tensorial group field theory: the melonic approximation,''
  Class.\ Quant.\ Grav.\  {\bf 32}, no. 17, 175012 (2015)
  doi:10.1088/0264-9381/32/17/175012
  [arXiv:1411.7213 [hep-th]].














\bibitem{Tetradis:1995br} 
  N.~Tetradis and D.~F.~Litim,
  ``Analytical solutions of exact renormalization group equations,''
  Nucl.\ Phys.\ B {\bf 464}, 492 (1996)
  doi:10.1016/0550-3213(95)00642-7
  [hep-th/9512073].




\bibitem{Litim:2000ci} 
  D.~F.~Litim,
  ``Optimization of the exact renormalization group,''
  Phys.\ Lett.\ B {\bf 486}, 92 (2000)
 doi:10.1016/S0370-2693(00)00748-6
  [hep-th/0005245].
  


\bibitem{Litim:2001dt} 
  D.~F.~Litim,
  ``Derivative expansion and renormalization group flows,''
  JHEP {\bf 0111}, 059 (2001)
  doi:10.1088/1126-6708/2001/11/059
  [hep-th/0111159].










\bibitem{Defenu:2014jfa} 
  N.~Defenu, P.~Mati, I.~G.~Marian, I.~Nandori and A.~Trombettoni,
  ``Truncation Effects in the Functional Renormalization Group Study of Spontaneous Symmetry Breaking,''
  JHEP {\bf 1505}, 141 (2015)
  doi:10.1007/JHEP05(2015)141
  [arXiv:1410.7024 [hep-th]].

\bibitem{Lahoche:2015ola} 
  V.~Lahoche, D.~Oriti and V.~Rivasseau,
  ``Renormalization of an Abelian Tensor Group Field Theory: Solution at Leading Order,''
  JHEP {\bf 1504}, 095 (2015)
  doi:10.1007/JHEP04(2015)095
  [arXiv:1501.02086 [hep-th]].


\bibitem{Rivasseau:2017xbk} 
  V.~Rivasseau and F.~Vignes-Tourneret,
  ``Constructive tensor field theory: The $T^{4}_{4}$ model,''
  arXiv:1703.06510 [math-ph].


\bibitem{Lionni:2016ush} 
  L.~Lionni and V.~Rivasseau,
  ``Intermediate Field Representation for Positive Matrix and Tensor Interactions,''
  arXiv:1609.05018 [math-ph].

\bibitem{Oriti:2013jga} 
  D.~Oriti,
  ``Disappearance and emergence of space and time in quantum gravity,''
  Stud.\ Hist.\ Phil.\ Sci.\ B {\bf 46}, 186 (2014)
  doi:10.1016/j.shpsb.2013.10.006
  [arXiv:1302.2849 [physics.hist-ph]].


\bibitem{Markopoulou:2007jf} 
  F.~Markopoulou,
  ``Conserved quantities in background independent theories,''
  J.\ Phys.\ Conf.\ Ser.\  {\bf 67}, 012019 (2007)
  doi:10.1088/1742-6596/67/1/012019
  [gr-qc/0703027].

\cite{Oriti:2013jga}-\cite{Wilkinson:2015fja}
\bibitem{Wilkinson:2015fja} 
  S.~A.~Wilkinson and A.~D.~Greentree,
  ``Geometrogenesis under Quantum Graphity: problems with the ripening Universe,''
  Phys.\ Rev.\ D {\bf 92}, no. 8, 084007 (2015)
  doi:10.1103/PhysRevD.92.084007
  [arXiv:1506.07588 [gr-qc]].

\end{thebibliography}
\end{document}